\documentclass[aps,prb,twocolumn, 10pt, superscriptaddress]{revtex4-2}
\usepackage{graphicx}
\usepackage{amsmath}
\usepackage{amssymb}
\usepackage{mathtools}
\usepackage{xcolor}
\usepackage[colorlinks=true, citecolor=blue, urlcolor=blue, linkcolor=blue,bookmarks=false,hypertexnames=true]{hyperref} 
\usepackage[T1]{fontenc} 
\usepackage{braket}

\begin{document}
\graphicspath{}

\title{Magnetism and hidden quantum geometry in charge neutral twisted trilayer graphene}

\author{Alina Wania Rodrigues} \thanks{awaniaro@uottawa.ca}
\affiliation{Department of Physics, University of Ottawa, Ottawa, Ontario, K1N 6N5, Canada}

\author{Maciej Bieniek}
\affiliation{Institute of Theoretical Physics, Wroc\l aw University of Science and Technology, Wybrze\.ze Wyspia\'nskiego 27, 50-370 Wroc\l aw, Poland}

\author{Daniel Miravet}
\affiliation{Department of Physics, University of Ottawa, Ottawa, Ontario, K1N 6N5, Canada}

\author{Pawe\l\ Hawrylak} \thanks{pawel.hawrylak@uottawa.ca}
\affiliation{Department of Physics, University of Ottawa, Ottawa, Ontario, K1N 6N5, Canada}

\date{\today}

\begin{abstract}
Here we present a theory of mirror-symmetric magic angle twisted trilayer graphene. The electronic properties are described by a Hubbard model with long range tunneling matrix elements. The electronic properties are obtained by solving the mean field Hubbard model. We obtain the bandstructure with characteristic flat bands and a  Dirac cone. At charge neutrality, turning on electron-electron interactions results in metallic to antiferomagnetic phase transition, for Hubbard interaction strength considerably smaller than in other graphene multilayers. We analyze the stability of the antiferromagnetic state against the symmetry breaking induced by hexagonal boron nitride encapsulation, and mirror symmetry breaking caused by the application of electric fields that mix the Dirac cone with the flat bands. Additionally, we explore the topological properties of the system, revealing a hidden quantum geometry. Despite the flat bands having zero Chern numbers, the multiband Berry curvature distribution over the moiré Brillouin zone exhibits a non-trivial structure. Finally, we propose a mechanism to tune this quantum geometry, providing a pathway to control the system's topological properties. 
\end{abstract}

\maketitle

\section*{Introduction}
Moir\'e materials offer a promising platform for designing tunable simulators of strongly correlated systems. The mismatch of the lattice constants or misalignment of the atomically-thin stacked layers leads to superlattice periodicity, which, for specific twist angles, flattens the bands around the Fermi energy. The flat bands enhance the role of the electron-electron interactions, leading to the observation of insulating and superconducting 
 phases \cite{Cao_Jarillo-Herrero_2018a, Cao_Jarillo-Herrero_2018b} for bilayer graphene twisted by the so-called magic angle \cite{Bistritzer_MacDonald_2011a}. Following this discovery, a significant number of moir\'e systems has been realized with different number of layers and sequences of twist angles, including, but not limited to, mirror-symmetric trilayers \cite{ Park_PJH_2021, Hao_Kim_2021, Cao_PJH_2021}, helical trilayers \cite{Devakul_Fu_2023}, double bilayers \cite{Su_Folk_2023} and pentalayers \cite{Han_Ju_2024}. Some of these have been shown to host a number of correlated and topological states such as correlated insulators \cite{Cao_Jarillo-Herrero_2018a, Tang_Fai_Mak_2020, Shen_Zhang_2020, Cao_PJH_2020, Liu_Kim_2020, Chen_Yankowitz_2021, Chen_Wang_2019}, quantum anomalous Hall effect \cite{Serlin_Young_2020}, ferromagnetism \cite{Sharpe_Goldhaber-Gordon_2019} and a generalized Wigner crystal state \cite{Regan_Wang_2020}. 

Mirror-symmetric magic angle twisted trilayer graphene (TTG) consists of a middle layer rotated by the magic angle relative to two aligned top and bottom layers \cite{Khalaf_Vishwanath_2019}. The band structure of TTG features two flat bands (per spin and valley) near the Fermi level, alongside Dirac cones that intersect the gap at the K points \cite{Khalaf_Vishwanath_2019, Carr_Kruchkov_2020, Calugaru_Bernevig_2021, Lei_MacDonald_2021, Xie_Lian_2021, Phong_Guinea_2021, Ledwith_Vishwanath_2021, Christos_Scheurer_2022, Yu_DasSarma_2023}. These features are also observed in ARPES experiments \cite{Li_Liu_2022}. The spectrum of TTG is highly tunable through several external parameters, including the electric field, the hexagonal boron nitride (hBN) alignment, and the twist angle. Compared to twisted bilayer graphene (TBG), TTG offers a highly stable and versatile platform for engineering electronic states. Notably, TTG demonstrates robust and reproducible superconductivity across a wide range of tuning parameters \cite{Park_PJH_2021, Cao_Jarillo-Herrero_2021, Hao_Kim_2021, Kim_Nadj-Perge_2022, Lin_Li_2022, Liu_Li_2022, Park_Jarillo-Herrero_2022, Siriviboon_Li_2022, Shen_Efetov_2023, Mukherjee_Deshmukh_2024, Batlle-Porro_Koppens_2024, Zhang_Li_2024, Zhou_Banerjee_2024}. Consequently, TTG has been the focus of an extensive theoretical investigation, employing both continuum models \cite{Khalaf_Vishwanath_2019, Calugaru_Bernevig_2021, Lei_MacDonald_2021, Xie_Lian_2021, Phong_Guinea_2021, Yu_DasSarma_2023} and atomistic approaches, such as \textit{ab initio} \cite{Carr_Kruchkov_2020, Fischer_Klebl_2022} and tight-binding \cite{Fischer_Klebl_2022}. Interaction effects in TTG have been explored in Refs. \cite{Xie_Lian_2021, Qin_MacDonald_2021, Christos_Scheurer_2022}, while theoretical descriptions of its superconducting phase are proposed in Refs. \cite{Chou_DasSarma_2021, Lake_Senthil_2021, Fischer_Klebl_2022, Gonzalez_Stauber_2023}. Additionally, the phonon properties of TTG were studied in Ref. \cite{Samajdar_Scheurer_2022}. Despite these significant advances, a unified theory of TTG is still lacking. Many of its properties, including the magnetic and topological phase diagrams, remain elusive.

In this work, we consider the atomistic tight-binding Hamiltonian combined with a Hubbard interaction term and apply the self-consistent Hartree-Fock method to determine the magnetic properties of the ground state. Our primary focus is on the transition from the paramagnetic to the antiferromagnetic (AF) phase, a phenomenon extensively studied in monolayer graphene systems \cite{Sorella_Tosatti_1992, Yazyev_2010, Guclu_Hawrylak_2014}. It has been shown that the area in the parameter space of the AF phase increases as we increase the number of layers \cite{Vahedi_TramblydeLaissardiere_2021}. We analyze the magnetic phase diagrams as functions of interaction strength, mirror, and sublattice symmetry-breaking perturbations, which arise from applied electric fields and hBN encapsulation. Interestingly, the interplay between magnetic ordering and topological properties, such as the emergence of AF Chern insulators \cite{Jiang_Wang_2018}, offers a compelling framework to explore quantum geometry in these systems. We compute the quasiparticle multiband Chern numbers for the valence and conduction flat bands, revealing the distribution of multiband Berry curvature (mBC) across the moiré Brillouin zone (mBZ), and demonstrate its tunability in experimentally realistic scenarios. We note that the behavior of mBC in TTG differs from the TBG case, where the $\Gamma$ point is the source of Berry's curvature \cite{Zhang_Lu_2024}. In the trilayer instance, we can identify contributions both from the $\Gamma$ and $K$ points.

\section*{Results}

\begin{figure}\
\includegraphics[width=0.5\textwidth]{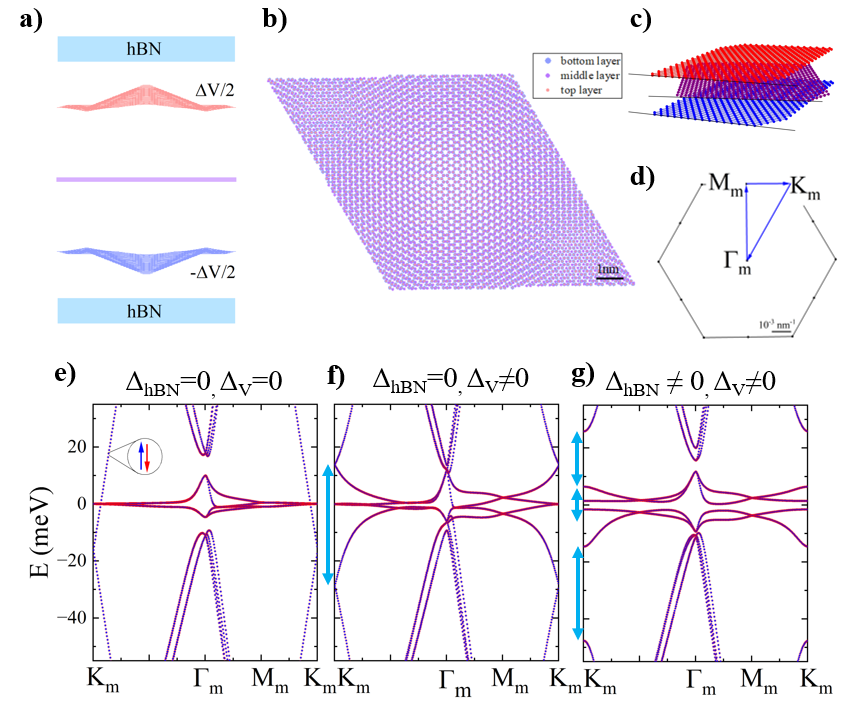}\
\caption{Structural and electronic properties of TTG. (a) Side, (b) top, and (c) 3D view of geometry. TTG consists of three layers of graphene, with the middle one twisted by $\theta$, with respect to the aligned top and bottom layers, which are relaxed out-of-plane. The system is encapsulated in aligned hBN, and a vertical electric field is applied. (d) mBZ with high symmetry points. (e) Band structure of pristine TTG. (f) Effect of non-zero electric field ($\Delta_{V} = 120$ meV) hybridizing Dirac cones with the flat band. (g) The combined effect of electric field and hBN encapsulation ($\Delta_{V}=120$meV$, \Delta_{\rm hBN}=25$meV), opening extra gaps near the Fermi level. The gaps opened by applying an electric field and hBN are marked with blue arrows.} \
\label{fig1}
\end{figure}\

\bfseries{Paramagnetic ground state, electric field- and hBN-induced gaps.} \normalfont We begin with an analysis of the paramagnetic ground state of the system. The single-particle Hamiltonian is given by:
\begin{equation}
    \hat{H}_{\rm TB} = \sum_{i,j} ^N \sum _\sigma t(\vec{r}_{i},\vec{r}_{j})(c_{i,\sigma}^{\dagger}c_{j,\sigma} + c_{j, \sigma}^{\dagger}c_{i, \sigma}),
\end{equation}
where hopping between $p_z$ orbitals in three layers can be modeled as
\begin{equation}
\begin{split}
    t(\vec{r}_{i},\vec{r}_{j}) = &(1-n^2)\gamma_0 \exp \left( \lambda_1 \left( 1 - \frac{|\vec{r}_i-\vec{r}_j|}{a} \right)\right)+ \\ 
    &n^2 \gamma_1 \exp \left(\lambda_2 \left(1-\frac{|\vec{r}_i-\vec{r}_j|}{c} \right) \right).
\end{split}
\label{hopping}
\end{equation}
Here $i,j$ are the indices enumerating atoms. We account for the electron-electron interactions through the Hubbard term:
\begin{equation}
\begin{split}
    \hat{H}_{U} = U\sum_{i}\hat{n}_{i\uparrow}\hat{n}_{i\downarrow}.
\end{split}
\label{hopping}
\end{equation}
The parameters and further details are given in Methods.

Figure \ref{fig1} (a-c) summarizes our system's geometry. The structure consists of top and bottom layers in AA stacking, with the middle layer twisted by an angle of $\theta = 1.55^\circ$. The TTG is encapsulated in aligned hBN, which is modeled as a staggered potential applied to the top and bottom layers: $\hat{H}_{\Delta_{\rm hBN}}=\Delta_{\textrm{hBN}} \sum_{i,\sigma} \alpha_i |l_i|  c_{i,\sigma}^\dagger c_{i,\sigma}$ where $l_i={1,0,-1}$ for top, middle and bottom layers respectively and $\alpha_i = {1,-1}$ when $i$ belongs to sublattice A and B respectively. This way there is a $2\Delta_{\textrm{hBN}}$ difference in onsite energy between A and B atoms of the top and bottom layers. Furthermore, a vertical electric field is introduced, modifying the potential difference between the top and bottom layers: $\hat{H}_{\Delta_{V}}=\frac{\Delta_{V}}{2} \sum_{i,\sigma} l_i c_{i,\sigma}^\dagger c_{i,\sigma}$. Therefore, the full Hamiltonian is given by 

\begin{equation}
    \hat{H} =  \hat{H}_{\rm TB} + \hat{H}_{\rm U}+\hat{H}_{\Delta_{\rm hBN}} + \hat{H}_{\Delta_{V}}.
    \label{eq:full_H}
\end{equation}

Since we assume a commensurate angle, i.e., periodic crystal structure, we can define a finite moir\'e unit cell (UC) (Fig. \ref{fig1}(b)) and mBZ (Fig. \ref{fig1}(d)). Each moir\'e UC consits of $N_{\rm at}$, and each atom repeats in the UC creating $N_{\rm at}$ simple Bravais sublattices. We apply periodic bondary condition to obtain allowed wavevectors, associate a Bloch wavefunction for a given wavevector $\vec{k}$ with each atom in a moir\'e cell, and diagonalize resulting Hamiltonian matrix to obtain energy bands. Fig.\ref{fig1}(e) shows the result for the interaction strength $U=0$. This procedure is described in greater detail in Ref \cite{Rodrigues_Hawrylak_2024}. To find the low energy state of this Hamiltonian we use a self consistent Hartree-Fock procedure, see details in Methods. The resulting ground state is paramagnetic, meaning each atom has an equal spin population (with precision of $10^{-6}$). We obtain the characteristic flat band at the Fermi level with two valence and two conduction bands for each spin component. In our calculation, we also reproduce the high velocity Dirac cones at moir\'e K points, with Dirac points approximately 15 meV below the Fermi level. We confirm that the wave functions of these Dirac cones are localized solely on the top and bottom layers, while the flat band's wave functions are localized $50\%$ on the middle layer and $25\%$ on the top and bottom (see Fig. 6 in SM). Such localization can be explained in terms of a TTG system comprising of a monolayer graphene and TBG \cite{Calugaru_Bernevig_2021}.

We now study the effect of vertical electric field and hBN encapsulation. We first introduce a non-zero electric field ($\Delta_{V}=120$meV). Such mirror-symmetry breaking perturbation hybridizes the Dirac cones with the flat bands \cite{Xie_Lian_2021}. This hybridization leads to splitting between the valence and conduction bands in the flat band, opening a gap at $K$ points, shown in Fig. \ref{fig1}(f) by a blue arrow. However, the gap at the Fermi level remains closed, and, in addition, the gap between the flat band and the remote bands closes at the $\Gamma$ point. 

In the next step, we turn on the interaction with encapsulating hBN in the form of a top and bottom layer staggered sublattice potential  Fig. \ref{fig1}(g) ($ \Delta_{\rm hBN}=25$meV). This leads to the opening of the gap within the flat band. Additionally, similar to the gapped monolayer graphene case \cite{Hunt_Ashoori_2013}, a gap between the Dirac cones opens and their dispersion becomes parabolic. The remote bands peel off from the flat band at the $\Gamma$ point so that our system consists of a well defined flat band and a gap around the Fermi energy. We note that TTG with an applied electric field and hBN is, therefore, strikingly similar to the TBG/hBN structure, with an extra set of remote bands at $K$ points originating from the gapped Dirac cones. More details of these mechanisms are available in SM.

\begin{figure*}\
\includegraphics[width=0.75\textwidth]{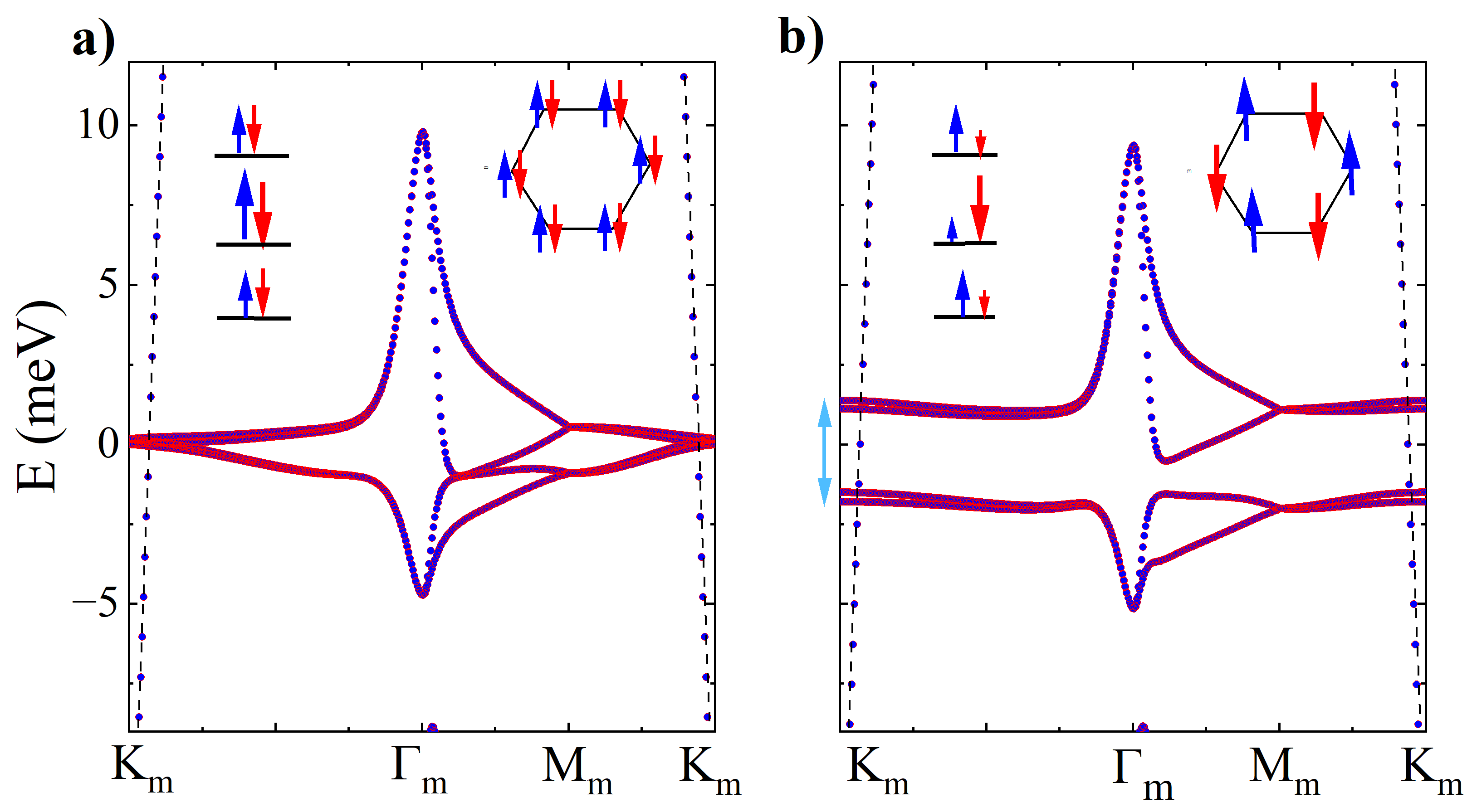}\
\caption{Hartree-Fock quasiparticle band structures of TTG with $\Delta_V=\Delta_{\rm hBN} = 0$. (a) Band structure in the paramagnetic state with $U=0.5t$. The insets show a schematic depiction of the spin configuration in the three layers (left) and on the honeycomb lattice of a single layer (right). Black dashed lines were added alongside the Dirac cone dispersion to improve readability. (b) Band structure of the flat band in the anti-ferromagnetic state with $U=1.1t$. Red and blue colors encode up and down spins, respectively. } \
\label{fig2}
\end{figure*}\

\begin{figure}\
\includegraphics[width=0.5\textwidth]{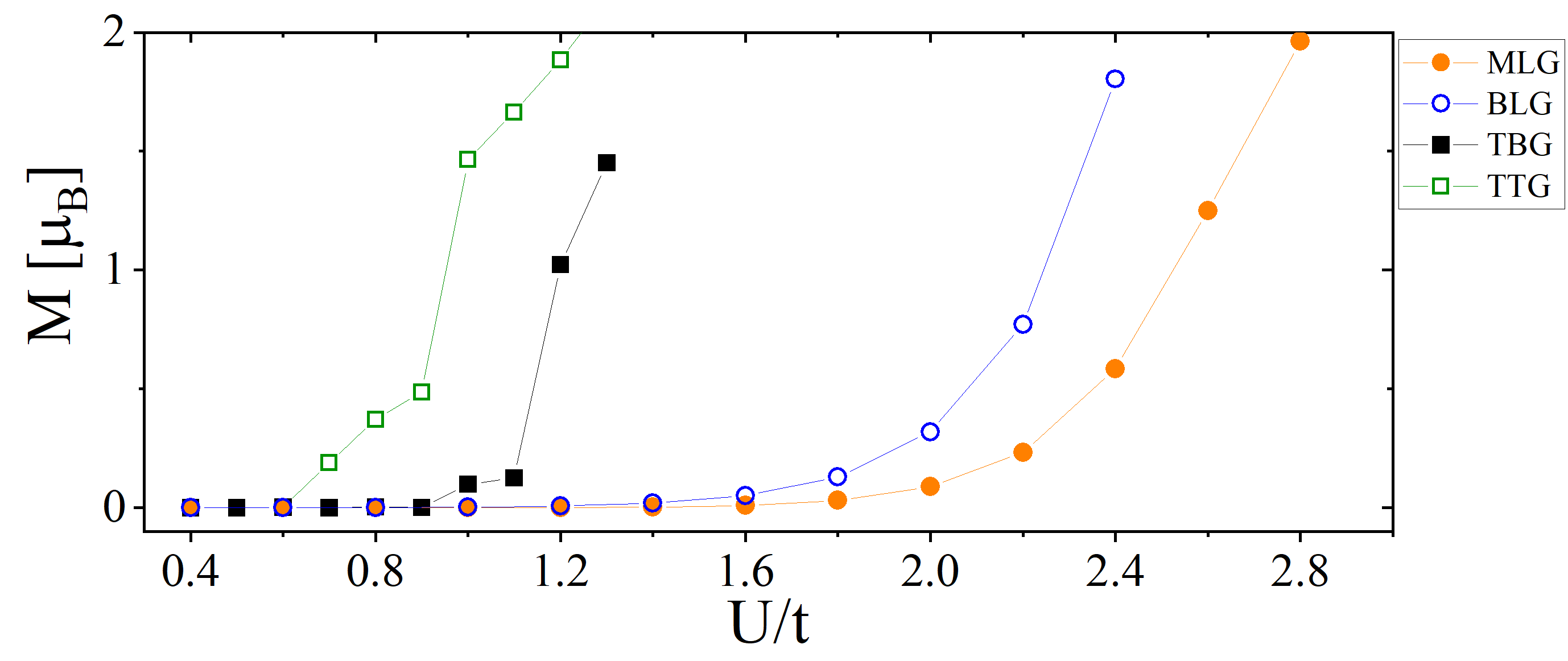}\
\caption{AF transition in various graphene-based systems. Total absolute magnetization of different graphene systems (MLG, BLG, TBG, TTG) in the units of Bohr magneton as a function of the interaction $U$ scaled by the nearest-neighbor hopping $t$. } \
\label{fig3}
\end{figure}\

\bfseries{Mott gap and anti-ferromagnetic transition.} \normalfont In the context of moir\'e materials, electron-electron interactions play an essential role due to the reduced bandwidth. To understand their influence, we consider different strengths of the interaction, modifying the Hubbard parameter $U$. We have tested several choices of initial input density and concluded that there is a phase transition between the non-magnetic and the AF state for the critical value of $U_{c}\approx0.6t$, where $t=|\gamma_0|=2.835$ eV is the nearest neighbor hopping. To determine the ground state of the system we analyze the total absolute magnetization (see Methods), which value is close to zero for a paramagnetic case, and significantly larger in the AF phase.  

We now show the Hartree-Fock quasiparticle bands for spin up and down obtained by self-consistently diagonalizing Hubbard Hamiltonian, Eq. \ref{eq:full_H}. For sufficiently small $U<U_{c}$, we do not observe any strong renormalization of the bands close to the Fermi level, as shown in Fig. \ref{fig2}(a). The state is paramagnetic, meaning each atom has an equal spin population. On the other hand, when $U>U_{c}$, a transition to an AF ground state occurs. A clear Mott gap is opened in the flat band after the phase transition in Fig. \ref{fig2}(b) for $U>1.1t$.  There is no gap opening for the Dirac cones, therefore, no gap is generated at the Fermi level. The AF state has the most significant spin polarization in the middle layer, which is the layer with the highest contribution to the flat band. Bilayer graphene has smaller $U_{c}$ than a monolayer, therefore the effect of $U$ is stronger in the middle of TTG. The total spin density on A and B sublattices of graphene layer building TTG is imbalanced, producing a microscopic AF state. Such AF state is analogous to an AF state in other n-layer graphene systems \cite{Vahedi_TramblydeLaissardiere_2021, Vidarte_Lewenkopf_2024}. We note that for the system to undergo a transition from a metallic to insulating state, the presence of an hBN substrate is necessary. Without it, the system is metallic both in the paramagnetic and antiferromagnetic states, while it remains insulating for all values of U and electric field with an hBN substrate present.

The Hubbard parameter $U$ for which the para- to anti-ferro- magnetic transition occurs is lower than in the mono- and bi-layer graphene systems (Fig. \ref{fig3}). For example, it has been shown that in the monolayer graphene (MLG) in the Hartree-Fock approximation $U_{c} \approx 2 t$ \cite{Vahedi_TramblydeLaissardiere_2021}. We reproduced this result in our model, with small differences caused by the long-range TB hopping necessary to model TTG correctly. Equivalent calculations for bilayer graphene (BLG) yield $U^{\rm BLG}_{c}<U^{\rm MLG}_{c}$. This stems from the fact that since the dispersion in BLG is parabolic, the density of states is increased, and the ratio of interactions to kinetic energy increases as well. It is not surprising then that TBG has a smaller critical transition parameter ($U=1.1 t$), as shown in Fig. \ref{fig3}. An extension of this Fig., showing the influence of the hBN substrate can be found in SM Fig. 7.

Naively, one could argue that since TTG has an additional Dirac cone and similar bandwidth of the flat bands, critical $U$ should be larger than in TBG. Our results point out that there is actually an opposite trend with $U^{\rm TTG}_{c} < U^{\rm TBG}_{c}$. This effect is a consequence of the TTG flat band being flatter than the TBG one. Here, we refer to its pieces being effectively flatter since, at the $\Gamma$ point, the flat band in TTG is actually wider (15 meV) than in TBG (8 meV). The flatter band enhances electron correlations, reducing screening effects and further decreasing the critical Hubbard parameters. 

We performed calculations for large U and concluded that the state with the lowest energy is still antiferromagnetic. Assuming ferromagnetic input (initial) state, we obtain energy difference between antiferromagnetic (ground) and ferromagnetic (higher in energy) states of $0.7193$ meV. This value is consistent in order of magnitude with TBG system, see e.g. recent works of Hou et al. \cite{Hou_Nevidomskyy_2025} and S\'anchez S\'anchez et al. \cite{Sanchez_Stauber_2025}, but at odds with Adhikari et al. \cite{Adhikari_Uchoa_2024} that claims obtaining ferromagnetic ground state at $\nu=0$ for TBG. \\

\begin{figure}\
\includegraphics[width=0.54\textwidth]{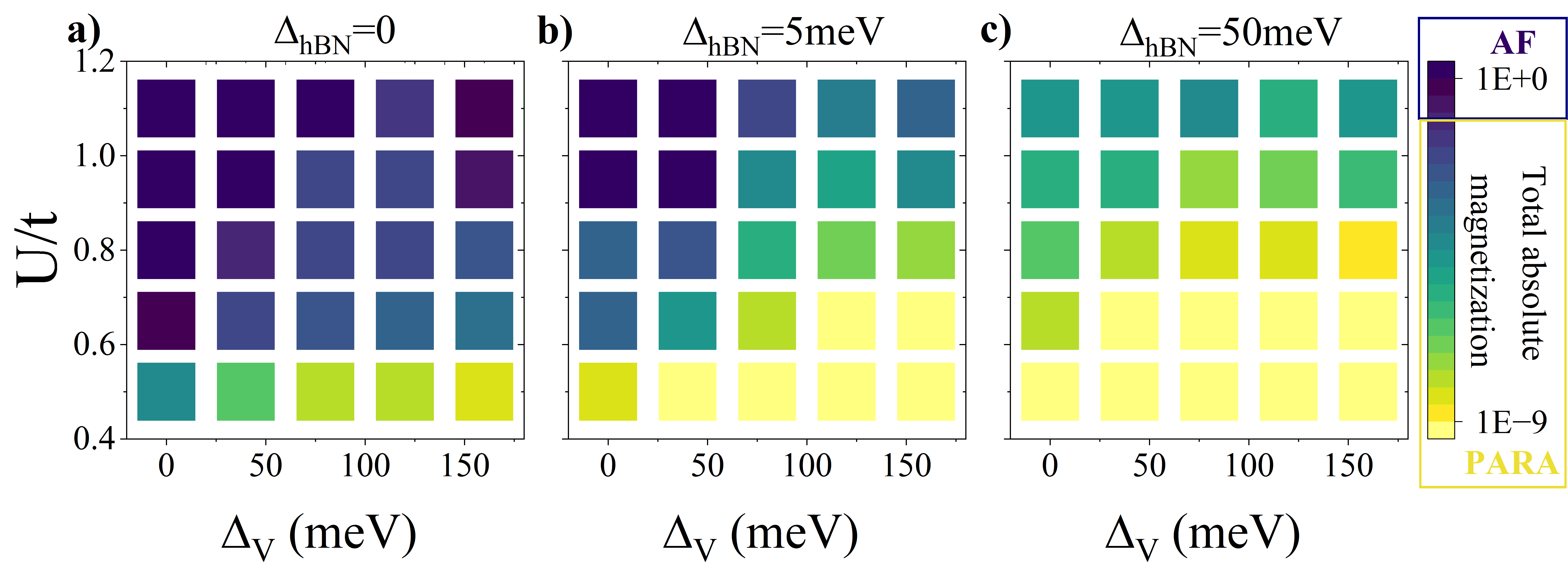}\
\caption{ Magnetic phase diagram of TTG. (a) TTG without a substrate. AF order parameter is studied in function of interaction strength (U/t) and applied electric potential ($\Delta_V$) (b-c) Similar phase diagram for staggered potential strengths (b) $\Delta_{hBN} = 5$ meV and (c) $\Delta_{hBN} = 50$ meV. The color scale denotes the total absolute magnetization in logarithmic scale. } \
\label{fig4}
\end{figure}\

\bfseries{Magnetic phase diagram stability.} \normalfont  We now study the combined effects of the applied electric field, hBN encapsulation, and the presence of electron-electron interactions on the ground state of charge neutral TTG. We study the magnetic phase diagram as a function of the interaction strength and electric field for three values of the staggered potential from hBN encapsulation. Our results are summarized in Fig. \ref{fig4}. Without the A/B sublattice symmetry breaking, we observe in Fig. \ref{fig4}(a) that the critical value of $U$ increases with increasing electric field. This follows from the fact that the applied electric field breaks the mirror-symmetry of TTG and causes the system to bear more resemblance to three monolayers, for which the critical value of $U/t$ is higher. Since in the actual device $U/t$ is fixed, the electric field allows thus to switch magnetic state from AF to paramagnetic for sufficiently large $U/t$. 

When small sublattice breaking perturbation is included in the top and bottom layers, even though most of the electron density is in the middle layer, a strong renormalization of the phase diagram is observed.
The imbalance introduced by the hBN layer directly counteracts the effects of the Hubbard term. While the Hubbard interaction penalizes the occupation of different-spin electrons on the same atom, the hBN-induced potential favors relocating electrons to sublattice B. Consequently, a larger value of $U$ is required to overcome this preference, transfer charge from sublattice B to sublattice A, and create a polarized state. The region of stable AF state is reduced to larger values of $U/t$ and smaller electric field; see top left of Fig. \ref{fig4}(b). Further increase of the staggered potential parameter, as in Fig. \ref{fig4}(c) to $\Delta_{\rm hBN}=50 $ meV, completely destroys AF region in the studied range of $U/t$. We confirm that the AF state is susceptible to mirror and sublattice symmetry-breaking perturbations.

\begin{figure}\
\includegraphics[width=0.5\textwidth]{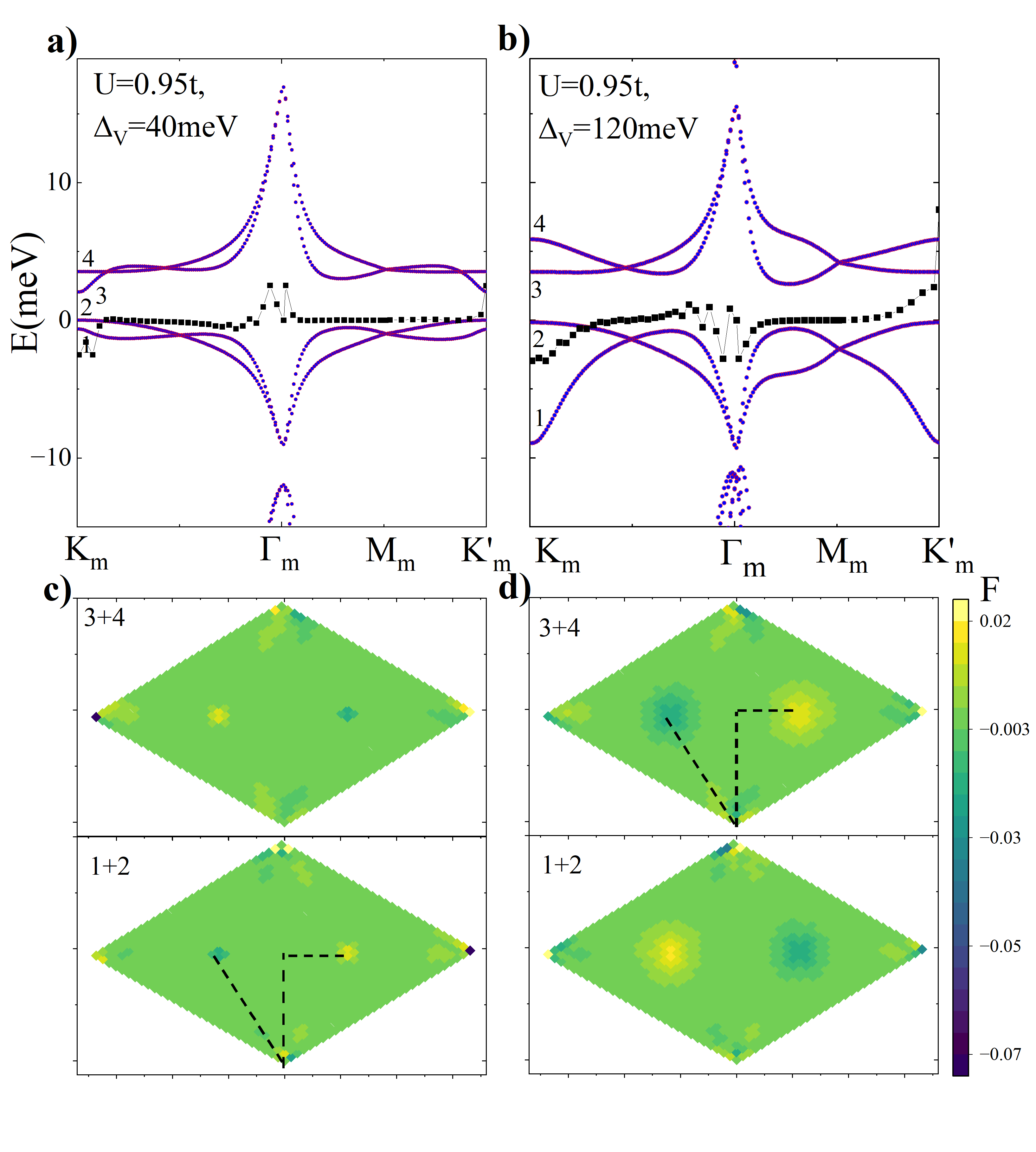}\
\caption{Multiband Berry's curvature in TTG. Band structure and mBC profiles along the mBZ path for both two VBs (1+2) and two CBs (3+4) for a) $\Delta_V = 40$ meV and b) $\Delta_V = 120$ meV. Distribution of mBC on whole mBZ c) corresponding to a) and d) corresponding to b). Color scale encodes the strength of mBC. The dashed line shows the mBZ path of a) and b). } \
\label{fig5}
\end{figure}\

\bfseries{Multiband Berry's curvature tuning.} \normalfont Flat bands of twisted materials are known to realize a variety of topological phases. For example, at $\nu=-3$ filling in TBG interaction-induced Chern insulator is predicted \cite{Sharpe_Goldhaber-Gordon_2019}. The topological phases catalog is still to be established in the TTG studied here. For a charge-neutral system for all studied electric fields, hBN strengths and Hubbard parameters $U$, flat bands near the Fermi level are always intertwined. Therefore, single band Berry's curvature and Chern numbers are not well defined. However, multiband analogs can be defined, in our case of 2+2 bands separated from other bands by well-defined energy gaps, valid for an ample region in the parameter space of $\Delta_{\textrm{hBN}}, \Delta_{V}$ and $U$. In Figs. \ref{fig5} (a) and (b), we show the corresponding Hartree-Fock band structures for two representative choices of parameters (more details in SM). Details of multiband Berry's curvature calculations are shown in the Methods section.

First, we establish that the multiband Chern number $C_{1+2+3+4}$ corresponding to all four bands is 0. This is also the case for the two flat valence (VBs) and conduction bands (CBs) separately ($C_{1+2}=C_{3+4}=0$). However, we observe a non-zero peak of mBC around $K$ and $K'$ points, see e.g., Fig \ref{fig5}(c), which cancel each other out. We establish that non-zero peaks are also present around the $\Gamma$ point, in Fig. \ref{fig5}(c) in the corners of rhomboidal moir\'e mBZ. In that sense, the geometry of wavefunctions is hidden since it produces $C=0$ despite having a rich profile in the mBZ. We note that significant Berry curvature occurring at moir\'e K points can be intuitively understood as resulting from the Dirac cone hybridizing with the flat-bands via electric field and hBN gap. This effect provides a new scenario for the generation of topological flat-bands in trilayer moir\'e materials.”

We also show that the k-space profiles of mBC can be tuned using an electric field. In Figs. \ref{fig5}(c) and (d), we compare the distribution of BC across mBZ for two values of $\Delta_V$, keeping fixed $\Delta_{\rm hBN}$ and $U$ for clarity. We observe that the spread of mBC increases for larger $\Delta_{V}$ around $K$ points. A less pronounced effect is observed around the $\Gamma$ point. Such curvature strength tuning, combined with the selective valley population for doped systems, might be interesting from the perspective of engineering interacting states. 

\section*{Discussion}

Twisted trilayer graphene is a highly tunable moir\'e system. We predict a surprisingly small critical value of Hubbard parameter $U$ necessary to observe magnetic phase transition between para- and anti-ferromagnetic states. This AF state is microscopic, in which spins are anti-aligned on neighboring graphene A and B sites inside TTG's AAA moir\'e centers. It is well-known that the AF state is not observed in pristine graphene because interactions are too weak. On the other hand, in TTG, this condition is weakened, although we note that such a state is destabilized by mirror and sublattice symmetry breaking perturbations. Effectively, we also expect stronger screening in our setup compared to that of suspended graphene, and further studies are necessary to establish optimal device design to reach the AF state in TTG. Such studies, including, e.g., quantum fluctuations, more realistic extended Hubbard models with long-range interaction, and better account for the dielectric environment and electron-electron screening models are the subject of our ongoing work. We note that the unrestricted Hartree-Fock method used in this paper has been shown sufficient to capture the ground state properties of twisted bilayer graphene \cite{Soejima_Zaletel_2020, Potasz_MacDonald_2021, Xie_Regnault_2021, Rai_Wehling_2024}, with general qualitative picture unchanged when more precise treatments of correlations, i.e. exact diagonalization, density matrix renormalization group or dynamical mean-field theory, are used. Inclusion of correlations is the subject of our future work. We note that in the absence of the hBN substrate, we did not observe the emergence of sublattice-polarized ground states. On the other hand, the detection of spontaneously broken valley symmetry \cite{Xie_Lian_2021, Christos_Scheurer_2022}, in the context of atomistic calculation, is not as straightforward, as shown in Ref. \cite{Sanchez_Stauber_2024}.

Our findings highlight the intricate interplay of electric field, sublattice symmetry breaking, and electron-electron interactions in determining the magnetic and electronic properties of TTG. The tunability of these parameters provides a promising pathway to designing devices with controllable correlated and topological phases. This study underscores TTG as a versatile platform to probe and manipulate quantum materials' magnetic and electronic properties.

Although topologically trivial at charge neutrality, TTG hosts nonzero peaks of multiband Berry's curvature. This curvature originates from the geometry of wave functions in the flat band and can be tuned using electric fields. This can be useful in understanding low electron/hole doped phases, especially those in which carriers unequally populate valleys, i.e., valley polarized states. It would also be interesting to establish if AF and topological phases can co-exist for realistic interactions in other multilayer systems, e.g., pentalayers with hBN-induced moir\'e potentials, in which fractional quantum anomalous Hall states have been recently observed \cite{Lu_Ju_2024}. 

\section*{Methods}

Mirror-symmetric TTG is constructed by aligning the top and bottom layers and twisting the middle layer by a given angle $\theta$. If $\theta$ belongs to a set of commensurate twist angles, moir\'e primitive vectors are $\vec{L}_1^{(m)} =m\vec{a}_1 - (m+1)\vec{a}_2$, $ \vec{L}_2^{(m)} =- (m+1)\vec{a}_1 - (2m+1)\vec{a}_2$, where $m$ is an integer, which is related to the twist angle  $\theta$ by $\cos(\theta_m) = (3m^2+3m+1/2)/(3m^2+3m+1)$. Here we use $\theta=1.55^{\circ}$ ($m=21$). The number of atoms in moir\'e unit cell is $N_{\rm at}=6(3m(m+1)+1)$, which here gives 8322 atoms. We account for the lattice relaxation effects along the z-axis by varying the inter-layer distance based on the stacking configuration \cite{Lopez-Bezanilla_Lado_2020, Carr_Kruchkov_2020, Fischer_Klebl_2022}. We treat the middle layer as rigid and relax only the top and bottom layers. More details of the relaxation are presented in SM. 

To describe the electronic properties of TTG, we employ a tight-binding model for $p_z$ atomic orbitals defined by Eq. (1) and (2). We use the following parameters $a=1.412$ \AA{}, $c=3.36$ \AA{}, $\gamma_0=-2.835$ eV, $\gamma_1 = 0.48$ eV. Here, $a$ is the carbon-carbon distance, $c$ is the inter-layer distance of an unrelaxed system, $\gamma_0 (\gamma_1)$ is the value of the nearest in-plane (out-of-plane) neighbor hopping. The dimensionless decay constants are $\lambda_1=3.15$ in-plane and $\lambda_2=7.50$ out-of-plane. A more detailed description of the model can be found in Refs. \cite{TramblydeLaissardiere_Magaud_2010, Rodrigues_Hawrylak_2024}. 

The considered values of the staggered potential induced by the presence of an hBN substrate were taken to be between $0-50$ meV. The order of these values is consistent with experiments \cite{Hunt_Ashoori_2013, Woods_Novoselov_2014} and \textit{ab initio} calculations \cite{Giovannetti_Brink_2007, Slawinska_Klusek_2010, Sachs_Lichtenstein_2011, Zollner_Fabian_2019, Zollner_Fabian_2021}  for different experimental set-ups, such as aligned graphene/hBN layers, entirely misaligned structure or modulated distance between TBG and the substrate.

The Hubbard repulsion term is given by $\hat{H}_U=U\sum_i \hat{n}_{i\uparrow}\hat{n}_{i\downarrow}$, where $\hat{n}_{i\sigma}=c_{i\sigma}^\dagger c_{i\sigma}$ is the spin-resolved electron number operator at site $i$. The parameter $U>0$ is the on-site Coulomb repulsion energy. We obtain the ground state of this system in the Hartree-Fock mean-field approximation, where $\hat{H}_U^{\rm MF}=U\sum_i \hat{n}_{i\uparrow}\langle \hat{n}_{i\downarrow} \rangle + \hat{n}_{i\downarrow}\langle \hat{n}_{i\uparrow} \rangle$. To study the magnetic properties of this ground state, we define the total, absolute magnetization in the units of Bohr magneton $\mu_B$ as $M= \sum_i^{N_{ \textrm{at} }} |m_{z,i}|=\sum_i^{N_{ \textrm{at} }}\frac{|\langle{\hat{n}_{i\uparrow}}\rangle-\langle{\hat{n}_{i\downarrow}}\rangle|}{2}$, where $m_{z,i}$ is the site magnetization.

We proceed numerically employing a self-consistent Hartree-Fock method. We perform an all-band calculation on a $8\times8$ grid in the momentum space until a difference of $10^{-8}$ is reached between the input and output density. We use several initial guesses, some of which exploit the system's symmetry, while some are generated randomly. We note that regardless of the initial condition, we obtain qualitatively the same converged density in all cases. Depending on the interaction strength, the system needs between 50-200 iterations to converge.

Finally, we calculate the multiband Chern numbers using a numerical procedure on a discretized Brillouin zone \cite{Fukui_Suzuki_2005}. We define $n$-bands link variable $U_\mu(\vec{k}_i) $ from the chosen set of wave functions $\psi=(\ket{\phi_1},...,\ket{\phi_n})$ as $U_\mu(\vec{k}_i) \equiv \det \psi^\dagger(\vec{k}_i) \psi(\vec{k}_i+\hat{\mu})/|\det \psi^\dagger(\vec{k}_i) \psi(\vec{k}_i+\hat{\mu})|$. Here, $\mu=1,2$ represents the possible directions in 2D while $\hat{\mu}$ is the vector connecting to the nearest $i$-th $k$ point with its nearest neighbor in direction $\mu$.   From the link variable, we can define mBZ as $F(\vec{k}_i)= \ln \left( U_{1}(\vec{k}_i) U_{2}(\vec{k}_i+\hat{1})U_{1}(\vec{k}_i+\hat{2})^{-1} U_{2}(\vec{k}_i)^{-1}\right)$. The  Chern number for the $n$ bands is then defined $c_{\psi} \equiv (2\pi i)^{-1} \sum_{i}F(\vec{k}_i)$. Note that since  $U_\mu(\vec{k}_i)$ is normalized, $F(\vec{k}_i)$ is purely imaginary, and the Chern number $c_{\psi}$  is real.

\section*{Data availability}
The data that support the findings of this study are available from the corresponding
author upon reasonable request.

\section*{Funding}
A.W.R., D.M. and P.H. were supported by NSERC Discovery Grant No. RGPIN 2019-05714, NSERC Alliance Quantum Grant No. ALLRP/578466-2022, the QSP-078 project of the Quantum Sensors Program at the National Research Council of Canada, University of Ottawa Research Chair in Quantum Theory of Materials, Nanostructures, and Devices. M.B. has been supported by the National Science Centre, Poland, under Grant No. 2021/43/D/ST3/01989. 
\section*{Acknowledgments}
This paper was partly enabled by support provided by the Digital Research Alliance of Canada \cite{AllianceCanada}. 

\bfseries{Author contributions} \normalfont A.W.R., M.B. and D.M. performed the numerical calculations.  A.W.R., M.B., D.M. and P.H. performed data analysis, discussed the results and wrote the manuscript with input from all co-authors.

\bibliography{bibliography_Dec2024}

@article{Cao_Jarillo-Herrero_2018a,
author={Cao, Yuan and Fatemi, Valla and Demir, Ahmet and Fang, Shiang and Tomarken, Spencer L. and Luo, Jason Y. and Sanchez-Yamagishi, Javier D. and Watanabe, Kenji and Taniguchi, Takashi and Kaxiras, Efthimios and Ashoori, Ray C. and Jarillo-Herrero, Pablo},
title={Correlated insulator behaviour at half-filling in magic-angle graphene superlattices},
journal={Nature},
year={2018},
month={Apr},
day={01},
volume={556},
number={7699},
pages={80-84},
issn={1476-4687},
doi={10.1038/nature26154},
url={https://doi.org/10.1038/nature26154}
}

@article{Cao_Jarillo-Herrero_2018b,
author={Cao, Yuan and Fatemi, Valla and Fang, Shiang and Watanabe, Kenji and Taniguchi, Takashi and Kaxiras, Efthimios and Jarillo-Herrero, Pablo},
title={Unconventional superconductivity in magic-angle graphene superlattices},
journal={Nature},
year={2018},
month={Apr},
day={01},
volume={556},
number={7699},
pages={43-50},
issn={1476-4687},
doi={10.1038/nature26160},
url={https://doi.org/10.1038/nature26160}
}

@article{TramblydeLaissardiere_Magaud_2010,
author = {Trambly de Laissardière, G. and Mayou, D. and Magaud, L.},
title = {Localization of Dirac Electrons in Rotated Graphene Bilayers},
journal = {Nano Letters},
volume = {10},
number = {3},
pages = {804-808},
year = {2010},
doi = {10.1021/nl902948m},
URL = {https://doi.org/10.1021/nl902948m}
}

@article{Vahedi_TramblydeLaissardiere_2021,
	title={{Magnetism of magic-angle twisted bilayer graphene}},
	author={Javad Vahedi and Robert Peters and Ahmed Missaoui and Andreas Honecker and Guy Trambly de Laissardiere},
	journal={SciPost Phys.},
	volume={11},
	pages={083},
	year={2021},
	publisher={SciPost},
	doi={10.21468/SciPostPhys.11.4.083},
	url={https://scipost.org/10.21468/SciPostPhys.11.4.083}
}

@article{Bistritzer_MacDonald_2011a,
	doi = {10.1073/pnas.1108174108},
	url = {https://doi.org/10.1073%2Fpnas.1108174108},
	year = 2011,
	month = {jul},
	publisher = {Proceedings of the National Academy of Sciences},
	volume = {108},
	number = {30},
	pages = {12233--12237},
	author = {Rafi Bistritzer and Allan H. MacDonald},
	title = {Moir{\'{e}
} bands in twisted double-layer graphene},
	journal = {Proceedings of the National Academy of Sciences}
}

@article{Cao_PJH_2021,
  author = {Cao, Yuan and Park, Jeong Min and Watanabe, Kenji and Taniguchi, Takashi and Jarillo-Herrero, Pablo},
  title = {Pauli-limit violation and re-entrant superconductivity in moiré graphene},
  journal = {Nature},
  year = {2021},
  volume = {595},
  number = {7868},
  pages = {526--531},
  doi = {10.1038/s41586-021-03685-y},
  issn = {1476-4687}
}

@article{Park_PJH_2021,
  author = {Park, Jeong Min and Cao, Yuan and Watanabe, Kenji and Taniguchi, Takashi and Jarillo-Herrero, Pablo},
  title = {Tunable strongly coupled superconductivity in magic-angle twisted trilayer graphene},
  journal = {Nature},
  year = {2021},
  volume = {590},
  number = {7845},
  pages = {249--255},
  doi = {10.1038/s41586-021-03192-0},
  issn = {1476-4687}
}

@article{Shen_Efetov_2023,
  author = {Shen, Cheng and Ledwith, Patrick J. and Watanabe, Kenji and Taniguchi, Takashi and Khalaf, Eslam and Vishwanath, Ashvin and Efetov, Dmitri K.},
  title = {Dirac spectroscopy of strongly correlated phases in twisted trilayer graphene},
  journal = {Nature Materials},
  year = {2023},
  volume = {22},
  number = {3},
  pages = {316--321},
  doi = {10.1038/s41563-022-01428-6},
  issn = {1476-4660}
}

@article{Devakul_Fu_2023,
  title = {Magic-angle helical trilayer graphene},
  author = {Devakul, Trithep and Ledwith, Patrick J. and Xia, Li-Qiao and Uri, Aviram and de la Barrera, Sergio C. and Jarillo-Herrero, Pablo and Fu, Liang},
  journal = {Science Advances},
  volume = {9},
  number = {36},
  pages = {eadi6063},
  year = {2023},
  publisher = {American Association for the Advancement of Science},
  doi = {10.1126/sciadv.adi6063},
  url = {https://doi.org/10.1126/sciadv.adi6063}
}

@article{Han_Ju_2024,
  author = {Han, Tonghang and Lu, Zhengguang and Scuri, Giovanni and Sung, Jiho and Wang, Jue and Han, Tianyi and Watanabe, Kenji and Taniguchi, Takashi and Park, Hongkun and Ju, Long},
  title = {Correlated insulator and {Chern} insulators in pentalayer rhombohedral-stacked graphene},
  journal = {Nature Nanotechnology},
  year = {2024},
  volume = {19},
  number = {2},
  pages = {181--187},
  doi = {10.1038/s41565-023-01520-1},
  issn = {1748-3395}
}

@article{Fukui_Suzuki_2005,
   title={Chern Numbers in Discretized {Brillouin} Zone: Efficient Method of Computing (Spin) {Hall} Conductances},
   volume={74},
   ISSN={1347-4073},
   url={http://dx.doi.org/10.1143/JPSJ.74.1674},
   DOI={10.1143/jpsj.74.1674},
   number={6},
   journal={Journal of the Physical Society of Japan},
   publisher={Physical Society of Japan},
   author={Fukui, Takahiro and Hatsugai, Yasuhiro and Suzuki, Hiroshi},
   year={2005},
   month=jun, pages={1674–1677} }

@article{Su_Folk_2023,
  author = {Su, Ruiheng and Kuiri, Manabendra and Watanabe, Kenji and Taniguchi, Takashi and Folk, Joshua},
  title = {Superconductivity in twisted double bilayer graphene stabilized by {WSe$_2$}},
  journal = {Nature Materials},
  year = {2023},
  volume = {22},
  number = {11},
  pages = {1332--1337},
  doi = {10.1038/s41563-023-01653-7},
  issn = {1476-4660}
}

@article{Tang_Fai_Mak_2020,
  author       = {Yanhao Tang and Lizhong Li and Tingxin Li and Yang Xu and Song Liu and Katayun Barmak and Kenji Watanabe and Takashi Taniguchi and Allan H. MacDonald and Jie Shan and Kin Fai Mak},
  title        = {Simulation of {Hubbard} model physics in {WSe$_2$/WS$_2$} moiré superlattices},
  journal      = {Nature},
  volume       = {579},
  number       = {7799},
  pages        = {353--358},
  year         = {2020},
  month        = {Mar},
  doi          = {10.1038/s41586-020-2085-3},
  url          = {https://doi.org/10.1038/s41586-020-2085-3},
  issn         = {1476-4687}
}

@article{Shen_Zhang_2020,
  author       = {Cheng Shen and Yanbang Chu and QuanSheng Wu and Na Li and Shuopei Wang and Yanchong Zhao and Jian Tang and Jieying Liu and Jinpeng Tian and Kenji Watanabe and Takashi Taniguchi and Rong Yang and Zi Yang Meng and Dongxia Shi and Oleg V. Yazyev and Guangyu Zhang},
  title        = {Correlated states in twisted double bilayer graphene},
  journal      = {Nature Physics},
  volume       = {16},
  number       = {5},
  pages        = {520--525},
  year         = {2020},
  month        = {May},
  doi          = {10.1038/s41567-020-0825-9},
  url          = {https://doi.org/10.1038/s41567-020-0825-9},
  issn         = {1745-2481}
}

@article{Regan_Wang_2020,
  author       = {Emma C. Regan and Danqing Wang and Chenhao Jin and M. Iqbal Bakti Utama and Beini Gao and Xin Wei and Sihan Zhao and Wenyu Zhao and Zuocheng Zhang and Kentaro Yumigeta and Mark Blei and Johan D. Carlström and Kenji Watanabe and Takashi Taniguchi and Sefaattin Tongay and Michael Crommie and Alex Zettl and Feng Wang},
  title        = {Mott and generalized {Wigner} crystal states in {WSe$_2$/WS$_2$} moiré superlattices},
  journal      = {Nature},
  volume       = {579},
  number       = {7799},
  pages        = {359--363},
  year         = {2020},
  month        = {Mar},
  doi          = {10.1038/s41586-020-2092-4},
  url          = {https://doi.org/10.1038/s41586-020-2092-4},
  issn         = {1476-4687}
}

@article{Cao_PJH_2020,
  author       = {Yuan Cao and Daniel Rodan-Legrain and Oriol Rubies-Bigorda and Jeong Min Park and Kenji Watanabe and Takashi Taniguchi and Pablo Jarillo-Herrero},
  title        = {Tunable correlated states and spin-polarized phases in twisted bilayer–bilayer graphene},
  journal      = {Nature},
  volume       = {583},
  number       = {7815},
  pages        = {215--220},
  year         = {2020},
  month        = {Jul},
  doi          = {10.1038/s41586-020-2260-6},
  url          = {https://doi.org/10.1038/s41586-020-2260-6},
  issn         = {1476-4687}
}

@article{Liu_Kim_2020,
  author       = {Xiaomeng Liu and Zeyu Hao and Eslam Khalaf and Jong Yeon Lee and Yuval Ronen and Hyobin Yoo and Danial Haei Najafabadi and Kenji Watanabe and Takashi Taniguchi and Ashvin Vishwanath and Philip Kim},
  title        = {Tunable spin-polarized correlated states in twisted double bilayer graphene},
  journal      = {Nature},
  volume       = {583},
  number       = {7815},
  pages        = {221--225},
  year         = {2020},
  month        = {Jul},
  doi          = {10.1038/s41586-020-2458-7},
  url          = {https://doi.org/10.1038/s41586-020-2458-7},
  issn         = {1476-4687}
}

@article{Chen_Yankowitz_2021,
  author       = {Shaowen Chen and Minhao He and Ya-Hui Zhang and Valerie Hsieh and Zaiyao Fei and K. Watanabe and T. Taniguchi and David H. Cobden and Xiaodong Xu and Cory R. Dean and Matthew Yankowitz},
  title        = {Electrically tunable correlated and topological states in twisted monolayer–bilayer graphene},
  journal      = {Nature Physics},
  volume       = {17},
  number       = {3},
  pages        = {374--380},
  year         = {2021},
  month        = {Mar},
  doi          = {10.1038/s41567-020-01062-6},
  url          = {https://doi.org/10.1038/s41567-020-01062-6},
  issn         = {1745-2481}
}

@article{Chen_Wang_2019,
  author       = {Guorui Chen and Lili Jiang and Shuang Wu and Bosai Lyu and Hongyuan Li and Bheema Lingam Chittari and Kenji Watanabe and Takashi Taniguchi and Zhiwen Shi and Jeil Jung and Yuanbo Zhang and Feng Wang},
  title        = {Evidence of a gate-tunable {Mott} insulator in a trilayer graphene moiré superlattice},
  journal      = {Nature Physics},
  volume       = {15},
  number       = {3},
  pages        = {237--241},
  year         = {2019},
  month        = {Mar},
  doi          = {10.1038/s41567-018-0387-2},
  url          = {https://doi.org/10.1038/s41567-018-0387-2},
  issn         = {1745-2481}
}

@article{Sharpe_Goldhaber-Gordon_2019,
  author       = {Sharpe, Aaron L. and Fox, Elijah J. and Barnard, Arthur W. and Finney, Joseph and Watanabe, Kenji and Taniguchi, Takashi and Kastner, Marc A. and Goldhaber-Gordon, David},
  title        = {Emergent ferromagnetism near three-quarters filling in twisted bilayer graphene},
  journal      = {Science},
  volume       = {365},
  number       = {6453},
  pages        = {605--608},
  year         = {2019},
  month        = {Aug},
  doi          = {10.1126/science.aaw3780},
  url          = {https://doi.org/10.1126/science.aaw3780},
  issn         = {1095-9203},
  pmid         = {31346139}
}

@article{Serlin_Young_2020,
  author       = {Serlin, Mark and Tschirhart, Charles L. and Polshyn, Hryhoriy and Zhang, Yaohui and Zhu, Jiamin and Watanabe, Kenji and Taniguchi, Takashi and Balents, Leon and Young, Andrea F.},
  title        = {Intrinsic quantized anomalous {Hall} effect in a moiré heterostructure},
  journal      = {Science},
  volume       = {367},
  number       = {6480},
  pages        = {900--903},
  year         = {2020},
  month        = {Feb},
  doi          = {10.1126/science.aay5533},
  url          = {https://doi.org/10.1126/science.aay5533},
  issn         = {1095-9203},
  pmid         = {31857492}
}

@article{Khalaf_Vishwanath_2019,
  title = {Magic angle hierarchy in twisted graphene multilayers},
  author = {Khalaf, Eslam and Kruchkov, Alex J. and Tarnopolsky, Grigory and Vishwanath, Ashvin},
  journal = {Phys. Rev. B},
  volume = {100},
  issue = {8},
  pages = {085109},
  numpages = {9},
  year = {2019},
  month = {Aug},
  publisher = {American Physical Society},
  doi = {10.1103/PhysRevB.100.085109},
  url = {https://link.aps.org/doi/10.1103/PhysRevB.100.085109}
}

@article{Calugaru_Bernevig_2021,
  title = {Twisted symmetric trilayer graphene: Single-particle and many-body {Hamiltonians} and hidden nonlocal symmetries of trilayer moir\'e systems with and without displacement field},
  author = {C\ifmmode \u{a}\else \u{a}\fi{}lug\ifmmode \u{a}\else \u{a}\fi{}ru, Dumitru and Xie, Fang and Song, Zhi-Da and Lian, Biao and Regnault, Nicolas and Bernevig, B. Andrei},
  journal = {Phys. Rev. B},
  volume = {103},
  issue = {19},
  pages = {195411},
  numpages = {45},
  year = {2021},
  month = {May},
  publisher = {American Physical Society},
  doi = {10.1103/PhysRevB.103.195411},
  url = {https://link.aps.org/doi/10.1103/PhysRevB.103.195411}
}

@article{Lei_MacDonald_2021,
  title = {Mirror symmetry breaking and lateral stacking shifts in twisted trilayer graphene},
  author = {Lei, Chao and Linhart, Lukas and Qin, Wei and Libisch, Florian and MacDonald, Allan H.},
  journal = {Phys. Rev. B},
  volume = {104},
  issue = {3},
  pages = {035139},
  numpages = {13},
  year = {2021},
  month = {Jul},
  publisher = {American Physical Society},
  doi = {10.1103/PhysRevB.104.035139},
  url = {https://link.aps.org/doi/10.1103/PhysRevB.104.035139}
}

@article{Xie_Lian_2021,
  title = {Twisted symmetric trilayer graphene. {II}. Projected {Hartree-Fock} study},
  author = {Xie, Fang and Regnault, Nicolas and C\ifmmode \u{a}\else \u{a}\fi{}lug\ifmmode \u{a}\else \u{a}\fi{}ru, Dumitru and Bernevig, B. Andrei and Lian, Biao},
  journal = {Phys. Rev. B},
  volume = {104},
  issue = {11},
  pages = {115167},
  numpages = {14},
  year = {2021},
  month = {Sep},
  publisher = {American Physical Society},
  doi = {10.1103/PhysRevB.104.115167},
  url = {https://link.aps.org/doi/10.1103/PhysRevB.104.115167}
}

@Article{Hao_Kim_2021,
author={Hao, Zeyu
and Zimmerman, A. M.
and Ledwith, Patrick
and Khalaf, Eslam
and Najafabadi, Danial Haie
and Watanabe, Kenji
and Taniguchi, Takashi
and Vishwanath, Ashvin
and Kim, Philip},
title={Electric field--tunable superconductivity in alternating-twist magic-angle trilayer graphene},
journal={Science},
year={2021},
month={Mar},
day={12},
publisher={American Association for the Advancement of Science},
volume={371},
number={6534},
pages={1133-1138},
doi={10.1126/science.abg0399},
url={https://doi.org/10.1126/science.abg0399}
}

@article{Phong_Guinea_2021,
  title = {Band structure and superconductivity in twisted trilayer graphene},
  author = {Phong, V\~o Tiến and Pantale\'on, Pierre A. and Cea, Tommaso and Guinea, Francisco},
  journal = {Phys. Rev. B},
  volume = {104},
  issue = {12},
  pages = {L121116},
  numpages = {6},
  year = {2021},
  month = {Sep},
  publisher = {American Physical Society},
  doi = {10.1103/PhysRevB.104.L121116},
  url = {https://link.aps.org/doi/10.1103/PhysRevB.104.L121116}
}

@article{Yu_DasSarma_2023,
  title = {Magic-angle twisted symmetric trilayer graphene as a topological heavy-fermion problem},
  author = {Yu, Jiabin and Xie, Ming and Bernevig, B. Andrei and Das Sarma, Sankar},
  journal = {Phys. Rev. B},
  volume = {108},
  issue = {3},
  pages = {035129},
  numpages = {54},
  year = {2023},
  month = {Jul},
  publisher = {American Physical Society},
  doi = {10.1103/PhysRevB.108.035129},
  url = {https://link.aps.org/doi/10.1103/PhysRevB.108.035129}
}

@article{Christos_Scheurer_2022,
  title = {Correlated Insulators, Semimetals, and Superconductivity in Twisted Trilayer Graphene},
  author = {Christos, Maine and Sachdev, Subir and Scheurer, Mathias S.},
  journal = {Phys. Rev. X},
  volume = {12},
  issue = {2},
  pages = {021018},
  numpages = {26},
  year = {2022},
  month = {Apr},
  publisher = {American Physical Society},
  doi = {10.1103/PhysRevX.12.021018},
  url = {https://link.aps.org/doi/10.1103/PhysRevX.12.021018}
}

@Article{Kim_Nadj-Perge_2022,
author={Kim, Hyunjin
and Choi, Youngjoon
and Lewandowski, Cyprian
and Thomson, Alex
and Zhang, Yiran
and Polski, Robert
and Watanabe, Kenji
and Taniguchi, Takashi
and Alicea, Jason
and Nadj-Perge, Stevan},
title={Evidence for unconventional superconductivity in twisted trilayer graphene},
journal={Nature},
year={2022},
month={Jun},
day={01},
volume={606},
number={7914},
pages={494-500},
issn={1476-4687},
doi={10.1038/s41586-022-04715-z},
url={https://doi.org/10.1038/s41586-022-04715-z}
}

@Article{Fischer_Klebl_2022,
author={Fischer, Ammon
and Goodwin, Zachary A. H.
and Mostofi, Arash A.
and Lischner, Johannes
and Kennes, Dante M.
and Klebl, Lennart},
title={Unconventional superconductivity in magic-angle twisted trilayer graphene},
journal={npj Quantum Materials},
year={2022},
month={Jan},
day={13},
volume={7},
number={1},
pages={5},
issn={2397-4648},
doi={10.1038/s41535-021-00410-w},
url={https://doi.org/10.1038/s41535-021-00410-w}
}

@article{Li_Liu_2022,
author = {Li, Yiwei and Zhang, Shihao and Chen, Fanqiang and Wei, Liyang and Zhang, Zonglin and Xiao, Hanbo and Gao, Han and Chen, Moyu and Liang, Shijun and Pei, Ding and Xu, Lixuan and Watanabe, Kenji and Taniguchi, Takashi and Yang, Lexian and Miao, Feng and Liu, Jianpeng and Cheng, Bin and Wang, Meixiao and Chen, Yulin and Liu, Zhongkai},
title = {Observation of Coexisting {Dirac} Bands and Moiré Flat Bands in Magic-Angle Twisted Trilayer Graphene},
journal = {Advanced Materials},
volume = {34},
number = {42},
pages = {2205996},
doi = {https://doi.org/10.1002/adma.202205996},
url = {https://onlinelibrary.wiley.com/doi/abs/10.1002/adma.202205996},
year = {2022}
}

@Article{Lin_Li_2022,
author={Lin, Jiang-Xiazi
and Siriviboon, Phum
and Scammell, Harley D.
and Liu, Song
and Rhodes, Daniel
and Watanabe, K.
and Taniguchi, T.
and Hone, James
and Scheurer, Mathias S.
and Li, J. I. A.},
title={Zero-field superconducting diode effect in small-twist-angle trilayer graphene},
journal={Nature Physics},
year={2022},
month={Oct},
day={01},
volume={18},
number={10},
pages={1221-1227},
issn={1745-2481},
doi={10.1038/s41567-022-01700-1},
url={https://doi.org/10.1038/s41567-022-01700-1}
}

@Article{Cao_Jarillo-Herrero_2021,
author={Cao, Yuan
and Park, Jeong Min
and Watanabe, Kenji
and Taniguchi, Takashi
and Jarillo-Herrero, Pablo},
title={Pauli-limit violation and re-entrant superconductivity in moir{\'e} graphene},
journal={Nature},
year={2021},
month={Jul},
day={01},
volume={595},
number={7868},
pages={526-531},
issn={1476-4687},
doi={10.1038/s41586-021-03685-y},
url={https://doi.org/10.1038/s41586-021-03685-y}
}

@Article{Liu_Li_2022,
author={Liu, Xiaoxue
and Zhang, Naiyuan James
and Watanabe, K.
and Taniguchi, T.
and Li, J. I. A.},
title={Isospin order in superconducting magic-angle twisted trilayer graphene},
journal={Nature Physics},
year={2022},
month={May},
day={01},
volume={18},
number={5},
pages={522-527},
issn={1745-2481},
doi={10.1038/s41567-022-01515-0},
url={https://doi.org/10.1038/s41567-022-01515-0}
}

@article{Chou_DasSarma_2021,
  title = {Correlation-Induced Triplet Pairing Superconductivity in Graphene-Based Moir\'e Systems},
  author = {Chou, Yang-Zhi and Wu, Fengcheng and Sau, Jay D. and Das Sarma, Sankar},
  journal = {Phys. Rev. Lett.},
  volume = {127},
  issue = {21},
  pages = {217001},
  numpages = {6},
  year = {2021},
  month = {Nov},
  publisher = {American Physical Society},
  doi = {10.1103/PhysRevLett.127.217001},
  url = {https://link.aps.org/doi/10.1103/PhysRevLett.127.217001}
}

@Article{Gonzalez_Stauber_2023,
author={Gonz{\'a}lez, J.
and Stauber, T.},
title={Ising superconductivity induced from spin-selective valley symmetry breaking in twisted trilayer graphene},
journal={Nature Communications},
year={2023},
month={May},
day={12},
volume={14},
number={1},
pages={2746},
issn={2041-1723},
doi={10.1038/s41467-023-38250-w},
url={https://doi.org/10.1038/s41467-023-38250-w}
}

@article{Lake_Senthil_2021,
  title = {Reentrant superconductivity through a quantum {Lifshitz} transition in twisted trilayer graphene},
  author = {Lake, Ethan and Senthil, T.},
  journal = {Phys. Rev. B},
  volume = {104},
  issue = {17},
  pages = {174505},
  numpages = {22},
  year = {2021},
  month = {Nov},
  publisher = {American Physical Society},
  doi = {10.1103/PhysRevB.104.174505},
  url = {https://link.aps.org/doi/10.1103/PhysRevB.104.174505}
}

@article{Ledwith_Vishwanath_2021,
      title={{TB} or not {TB}? {Contrasting} properties of twisted bilayer graphene and the alternating twist $n$-layer structures ($n=3, 4, 5, \dots$)}, 
      author={Patrick J. Ledwith and Eslam Khalaf and Ziyan Zhu and Stephen Carr and Efthimios Kaxiras and Ashvin Vishwanath},
      year={2021},
      eprint={2111.11060},
      journal={arXiv},
      primaryClass={cond-mat.str-el},
      url={https://arxiv.org/abs/2111.11060}, 
}

@article{Mukherjee_Deshmukh_2024,
      title={Superconducting magic-angle twisted trilayer graphene hosts competing magnetic order and moir\'e inhomogeneities}, 
      author={Ayshi Mukherjee and Surat Layek and Subhajit Sinha and Ritajit Kundu and Alisha H. Marchawala and Mahesh Hingankar and Joydip Sarkar and L. D. Varma Sangani and Heena Agarwal and Sanat Ghosh and Aya Batoul Tazi and Kenji Watanabe and Takashi Taniguchi and Abhay N. Pasupathy and Arijit Kundu and Mandar M. Deshmukh},
      year={2024},
      eprint={2406.02521},
      journal={arXiv},
      primaryClass={cond-mat.mes-hall},
      url={https://arxiv.org/abs/2406.02521}, 
}

@Article{Park_Jarillo-Herrero_2022,
author={Park, Jeong Min
and Cao, Yuan
and Xia, Li-Qiao
and Sun, Shuwen
and Watanabe, Kenji
and Taniguchi, Takashi
and Jarillo-Herrero, Pablo},
title={Robust superconductivity in magic-angle multilayer graphene family},
journal={Nature Materials},
year={2022},
month={Aug},
day={01},
volume={21},
number={8},
pages={877-883},
issn={1476-4660},
doi={10.1038/s41563-022-01287-1},
url={https://doi.org/10.1038/s41563-022-01287-1}
}

@article{Batlle-Porro_Koppens_2024,
      title={Cryo-Near-Field Photovoltage Microscopy of Heavy-Fermion Twisted Symmetric Trilayer Graphene}, 
      author={Sergi Batlle-Porro and Dumitru Calugaru and Haoyu Hu and Roshan Krishna Kumar and Niels C. H. Hesp and Kenji Watanabe and Takashi Taniguchi and B. Andrei Bernevig and Petr Stepanov and Frank H. L. Koppens},
      year={2024},
      eprint={2402.12296},
      journal={arXiv},
      primaryClass={cond-mat.mes-hall},
      url={https://arxiv.org/abs/2402.12296}, 
}

@article{Qin_MacDonald_2021,
  title = {In-Plane Critical Magnetic Fields in Magic-Angle Twisted Trilayer Graphene},
  author = {Qin, Wei and MacDonald, Allan H.},
  journal = {Phys. Rev. Lett.},
  volume = {127},
  issue = {9},
  pages = {097001},
  numpages = {7},
  year = {2021},
  month = {Aug},
  publisher = {American Physical Society},
  doi = {10.1103/PhysRevLett.127.097001},
  url = {https://link.aps.org/doi/10.1103/PhysRevLett.127.097001}
}

@article{Samajdar_Scheurer_2022,
  title = {Moir\'e phonons and impact of electronic symmetry breaking in twisted trilayer graphene},
  author = {Samajdar, Rhine and Teng, Yanting and Scheurer, Mathias S.},
  journal = {Phys. Rev. B},
  volume = {106},
  issue = {20},
  pages = {L201403},
  numpages = {6},
  year = {2022},
  month = {Nov},
  publisher = {American Physical Society},
  doi = {10.1103/PhysRevB.106.L201403},
  url = {https://link.aps.org/doi/10.1103/PhysRevB.106.L201403}
}

@article{Siriviboon_Li_2022,
      title={A new flavor of correlation and superconductivity in small twist-angle trilayer graphene}, 
      author={Phum Siriviboon and Jiang-Xiazi Lin and Xiaoxue Liu and Harley D. Scammell and Song Liu and Daniel Rhodes and K. Watanabe and T. Taniguchi and James Hone and Mathias S. Scheurer and J. I. A. Li},
      year={2022},
      eprint={2112.07127},
      journal={arXiv},
      primaryClass={cond-mat.mes-hall},
      url={https://arxiv.org/abs/2112.07127}, 
}

@Article{Zhang_Li_2024,
author={Zhang, Naiyuan James
and Lin, Jiang-Xiazi
and Chichinadze, Dmitry V.
and Wang, Yibang
and Watanabe, Kenji
and Taniguchi, Takashi
and Fu, Liang
and Li, J. I. A.},
title={Angle-resolved transport non-reciprocity and spontaneous symmetry breaking in twisted trilayer graphene},
journal={Nature Materials},
year={2024},
month={Mar},
day={01},
volume={23},
number={3},
pages={356-362},
issn={1476-4660},
doi={10.1038/s41563-024-01809-z},
url={https://doi.org/10.1038/s41563-024-01809-z}
}

@article{Zhou_Banerjee_2024,
      title={Double-dome Unconventional Superconductivity in Twisted Trilayer Graphene}, 
      author={Zekang Zhou and Jin Jiang and Paritosh Karnatak and Ziwei Wang and Glenn Wagner and Kenji Watanabe and Takashi Taniguchi and Christian Schönenberger and S. A. Parameswaran and Steven H. Simon and Mitali Banerjee},
      year={2024},
      eprint={2404.09909},
      journal={arXiv},
      primaryClass={cond-mat.mes-hall},
      url={https://arxiv.org/abs/2404.09909}, 
}

@article{Lopez-Bezanilla_Lado_2020,
  title = {Electrical band flattening, valley flux, and superconductivity in twisted trilayer graphene},
  author = {Lopez-Bezanilla, Alejandro and Lado, J. L.},
  journal = {Phys. Rev. Res.},
  volume = {2},
  issue = {3},
  pages = {033357},
  numpages = {10},
  year = {2020},
  month = {Sep},
  publisher = {American Physical Society},
  doi = {10.1103/PhysRevResearch.2.033357},
  url = {https://link.aps.org/doi/10.1103/PhysRevResearch.2.033357}
}

@Article{Carr_Kruchkov_2020,
author={Carr, Stephen
and Li, Chenyuan
and Zhu, Ziyan
and Kaxiras, Efthimios
and Sachdev, Subir
and Kruchkov, Alexander},
title={Ultraheavy and Ultrarelativistic {Dirac} Quasiparticles in Sandwiched Graphenes},
journal={Nano Letters},
year={2020},
month={May},
day={13},
publisher={American Chemical Society},
volume={20},
number={5},
pages={3030-3038},
issn={1530-6984},
doi={10.1021/acs.nanolett.9b04979},
url={https://doi.org/10.1021/acs.nanolett.9b04979}
}

@article{Sorella_Tosatti_1992,
doi = {10.1209/0295-5075/19/8/007},
url = {https://dx.doi.org/10.1209/0295-5075/19/8/007},
year = {1992},
month = {aug},
publisher = {},
volume = {19},
number = {8},
pages = {699},
author = {S. Sorella and  E. Tosatti},
title = {Semi-Metal-Insulator Transition of the {Hubbard} Model in the Honeycomb Lattice},
journal = {Europhysics Letters}
}

@article{Yazyev_2010,
doi = {10.1088/0034-4885/73/5/056501},
url = {https://dx.doi.org/10.1088/0034-4885/73/5/056501},
year = {2010},
month = {apr},
publisher = {},
volume = {73},
number = {5},
pages = {056501},
author = {Oleg V Yazyev},
title = {Emergence of magnetism in graphene materials and nanostructures},
journal = {Reports on Progress in Physics}
}

@Inbook{Guclu_Hawrylak_2014,
author="G{\"u}{\c{c}}l{\"u}, Alev Devrim
and Potasz, Pawel
and Korkusinski, Marek
and Hawrylak, Pawel",
title="Magnetic Properties of Gated Graphene Nanostructures",
bookTitle="Graphene Quantum Dots",
year="2014",
publisher="Springer Berlin Heidelberg",
address="Berlin, Heidelberg",
pages="111--144",
isbn="978-3-662-44611-9",
doi="10.1007/978-3-662-44611-9_6",
url="https://doi.org/10.1007/978-3-662-44611-9_6"
}

@Article{Lu_Ju_2024,
author={Lu, Zhengguang
and Han, Tonghang
and Yao, Yuxuan
and Reddy, Aidan P.
and Yang, Jixiang
and Seo, Junseok
and Watanabe, Kenji
and Taniguchi, Takashi
and Fu, Liang
and Ju, Long},
title={Fractional quantum anomalous {Hall} effect in multilayer graphene},
journal={Nature},
year={2024},
month={Feb},
day={01},
volume={626},
number={8000},
pages={759-764},
issn={1476-4687},
doi={10.1038/s41586-023-07010-7},
url={https://doi.org/10.1038/s41586-023-07010-7}
}

@article{Jiang_Wang_2018,
  title = {Antiferromagnetic {Chern} Insulators in Noncentrosymmetric Systems},
  author = {Jiang, Kun and Zhou, Sen and Dai, Xi and Wang, Ziqiang},
  journal = {Phys. Rev. Lett.},
  volume = {120},
  issue = {15},
  pages = {157205},
  numpages = {5},
  year = {2018},
  month = {Apr},
  publisher = {American Physical Society},
  doi = {10.1103/PhysRevLett.120.157205},
  url = {https://link.aps.org/doi/10.1103/PhysRevLett.120.157205}
}

@article{Zhang_Lu_2024,
      title={Commensurate and Incommensurate {Chern} Insulators in Magic-angle Bilayer Graphene}, 
      author={Zaizhe Zhang and Jingxin Yang and Bo Xie and Zuo Feng and Shu Zhang and Kenji Watanabe and Takashi Taniguchi and Xiaoxia Yang and Qing Dai and Tao Liu and Donghua Liu and Kaihui Liu and Zhida Song and Jianpeng Liu and Xiaobo Lu},
      year={2024},
      eprint={2408.12509},
      journal={arXiv},
      primaryClass={cond-mat.mes-hall},
      url={https://arxiv.org/abs/2408.12509}, 
}

@article{Vidarte_Lewenkopf_2024,
      title={Magnetic properties of low-angle twisted bilayer graphene at three-quarters filling}, 
      author={Kevin J. U. Vidarte and Caio Lewenkopf},
      year={2024},
      eprint={2404.08177},
      journal={arXiv},
      primaryClass={cond-mat.mes-hall},
      url={https://arxiv.org/abs/2404.08177}, 
}

@misc{AllianceCanada,
  author = {{Digital Research Alliance of Canada}},
  title = {Digital Research Alliance of Canada},
  howpublished = {\url{www.alliancecan.ca}}
}

@article{Rodrigues_Hawrylak_2024,
  title = {Atomistic theory of the moir\'e {Hofstadter} butterfly in magic-angle graphene},
  author = {Wania Rodrigues, Alina and Bieniek, Maciej and Potasz, Pawe\l{} and Miravet, Daniel and Thomale, Ronny and Korkusi\ifmmode \acute{n}\else \'{n}\fi{}ski, Marek and Hawrylak, Pawe\l{}},
  journal = {Phys. Rev. B},
  volume = {109},
  issue = {7},
  pages = {075166},
  numpages = {18},
  year = {2024},
  month = {Feb},
  publisher = {American Physical Society},
  doi = {10.1103/PhysRevB.109.075166},
  url = {https://link.aps.org/doi/10.1103/PhysRevB.109.075166}
}

@article{Hunt_Ashoori_2013,
author = {B. Hunt  and J. D. Sanchez-Yamagishi  and A. F. Young  and M. Yankowitz  and B. J. LeRoy  and K. Watanabe  and T. Taniguchi  and P. Moon  and M. Koshino  and P. Jarillo-Herrero  and R. C. Ashoori },
title = {{Massive Dirac Fermions and Hofstadter Butterfly in a van der Waals Heterostructure}},
journal = {Science},
volume = {340},
number = {6139},
pages = {1427-1430},
year = {2013},
doi = {10.1126/science.1237240},
URL = {https://www.science.org/doi/abs/10.1126/science.1237240}
}

@article{Soejima_Zaletel_2020,
  title = {Efficient simulation of moir\'e materials using the density matrix renormalization group},
  author = {Soejima, Tomohiro and Parker, Daniel E. and Bultinck, Nick and Hauschild, Johannes and Zaletel, Michael P.},
  journal = {Phys. Rev. B},
  volume = {102},
  issue = {20},
  pages = {205111},
  numpages = {26},
  year = {2020},
  month = {Nov},
  publisher = {American Physical Society},
  doi = {10.1103/PhysRevB.102.205111},
  url = {https://link.aps.org/doi/10.1103/PhysRevB.102.205111}
}

@article{Potasz_MacDonald_2021,
  title = {Exact Diagonalization for Magic-Angle Twisted Bilayer Graphene},
  author = {Potasz, Pawel and Xie, Ming and MacDonald, A. H.},
  journal = {Phys. Rev. Lett.},
  volume = {127},
  issue = {14},
  pages = {147203},
  numpages = {6},
  year = {2021},
  month = {Sep},
  publisher = {American Physical Society},
  doi = {10.1103/PhysRevLett.127.147203},
  url = {https://link.aps.org/doi/10.1103/PhysRevLett.127.147203}
}

@article{Xie_Regnault_2021,
  title = {Twisted bilayer graphene. VI. An exact diagonalization study at nonzero integer filling},
  author = {Xie, Fang and Cowsik, Aditya and Song, Zhi-Da and Lian, Biao and Bernevig, B. Andrei and Regnault, Nicolas},
  journal = {Phys. Rev. B},
  volume = {103},
  issue = {20},
  pages = {205416},
  numpages = {41},
  year = {2021},
  month = {May},
  publisher = {American Physical Society},
  doi = {10.1103/PhysRevB.103.205416},
  url = {https://link.aps.org/doi/10.1103/PhysRevB.103.205416}
}

@article{Rai_Wehling_2024,
  title = {Dynamical Correlations and Order in Magic-Angle Twisted Bilayer Graphene},
  author = {Rai, Gautam and Crippa, Lorenzo and C\ifmmode \u{a}\else \u{a}\fi{}lug\ifmmode \u{a}\else \u{a}\fi{}ru, Dumitru and Hu, Haoyu and Paoletti, Francesca and de' Medici, Luca and Georges, Antoine and Bernevig, B. Andrei and Valent\'{\i}, Roser and Sangiovanni, Giorgio and Wehling, Tim},
  journal = {Phys. Rev. X},
  volume = {14},
  issue = {3},
  pages = {031045},
  numpages = {22},
  year = {2024},
  month = {Sep},
  publisher = {American Physical Society},
  doi = {10.1103/PhysRevX.14.031045},
  url = {https://link.aps.org/doi/10.1103/PhysRevX.14.031045}
}

@article{Hou_Nevidomskyy_2025,
  title = {Particle-hole asymmetric phases in doped twisted bilayer graphene},
  author = {Hou, Run and Sur, Shouvik and Wagner, Lucas K. and Nevidomskyy, Andriy H.},
  journal = {Phys. Rev. B},
  volume = {111},
  issue = {12},
  pages = {125140},
  numpages = {20},
  year = {2025},
  month = {Mar},
  publisher = {American Physical Society},
  doi = {10.1103/PhysRevB.111.125140},
  url = {https://link.aps.org/doi/10.1103/PhysRevB.111.125140}
}

@article{Sanchez_Stauber_2025,
  title = {Nonflat bands and chiral symmetry in magic-angle twisted bilayer graphene},
  author = {S\'anchez S\'anchez, Miguel and Gonz\'alez, Jos\'e and Stauber, Tobias},
  journal = {Phys. Rev. B},
  volume = {111},
  issue = {20},
  pages = {205133},
  numpages = {19},
  year = {2025},
  month = {May},
  publisher = {American Physical Society},
  doi = {10.1103/PhysRevB.111.205133},
  url = {https://link.aps.org/doi/10.1103/PhysRevB.111.205133}
}

@article{Adhikari_Uchoa_2024,
  title = {Strongly interacting phases in twisted bilayer graphene at the magic angle},
  author = {Adhikari, Khagendra and Seo, Kangjun and Beach, K. S. D. and Uchoa, Bruno},
  journal = {Phys. Rev. B},
  volume = {110},
  issue = {12},
  pages = {L121123},
  numpages = {7},
  year = {2024},
  month = {Sep},
  publisher = {American Physical Society},
  doi = {10.1103/PhysRevB.110.L121123},
  url = {https://link.aps.org/doi/10.1103/PhysRevB.110.L121123}
}

@article{Woods_Novoselov_2014,
  author       = {C. R. Woods and L. Britnell and A. Eckmann and R. S. Ma and J. C. Lu and H. M. Guo and X. Lin and G. L. Yu and Y. Cao and R. V. Gorbachev and A. V. Kretinin and J. Park and L. A. Ponomarenko and M. I. Katsnelson and Yu. N. Gornostyrev and K. Watanabe and T. Taniguchi and C. Casiraghi and H.-J. Gao and A. K. Geim and K. S. Novoselov},
  title        = {Commensurate–incommensurate transition in graphene on hexagonal boron nitride},
  journal      = {Nature Physics},
  year         = {2014},
  volume       = {10},
  number       = {6},
  pages        = {451--456},
  doi          = {10.1038/nphys2954},
  url          = {https://doi.org/10.1038/nphys2954}
}

@article{Giovannetti_Brink_2007,
  title = {Substrate-induced band gap in graphene on hexagonal boron nitride: Ab initio density functional calculations},
  author = {Giovannetti, Gianluca and Khomyakov, Petr A. and Brocks, Geert and Kelly, Paul J. and van den Brink, Jeroen},
  journal = {Phys. Rev. B},
  volume = {76},
  issue = {7},
  pages = {073103},
  numpages = {4},
  year = {2007},
  month = {Aug},
  publisher = {American Physical Society},
  doi = {10.1103/PhysRevB.76.073103},
  url = {https://link.aps.org/doi/10.1103/PhysRevB.76.073103}
}

@article{Slawinska_Klusek_2010,
  title = {Energy gap tuning in graphene on hexagonal boron nitride bilayer system},
  author = {S\l{}awi\ifmmode \acute{n}\else \'{n}\fi{}ska, J. and Zasada, I. and Klusek, Z.},
  journal = {Phys. Rev. B},
  volume = {81},
  issue = {15},
  pages = {155433},
  numpages = {9},
  year = {2010},
  month = {Apr},
  publisher = {American Physical Society},
  doi = {10.1103/PhysRevB.81.155433},
  url = {https://link.aps.org/doi/10.1103/PhysRevB.81.155433}
}

@article{Sachs_Lichtenstein_2011,
  title = {Adhesion and electronic structure of graphene on hexagonal boron nitride substrates},
  author = {Sachs, B. and Wehling, T. O. and Katsnelson, M. I. and Lichtenstein, A. I.},
  journal = {Phys. Rev. B},
  volume = {84},
  issue = {19},
  pages = {195414},
  numpages = {7},
  year = {2011},
  month = {Nov},
  publisher = {American Physical Society},
  doi = {10.1103/PhysRevB.84.195414},
  url = {https://link.aps.org/doi/10.1103/PhysRevB.84.195414}
}

@article{Zollner_Fabian_2019,
  title = {Heterostructures of graphene and hBN: Electronic, spin-orbit, and spin relaxation properties from first principles},
  author = {Zollner, Klaus and Gmitra, Martin and Fabian, Jaroslav},
  journal = {Phys. Rev. B},
  volume = {99},
  issue = {12},
  pages = {125151},
  numpages = {16},
  year = {2019},
  month = {Mar},
  publisher = {American Physical Society},
  doi = {10.1103/PhysRevB.99.125151},
  url = {https://link.aps.org/doi/10.1103/PhysRevB.99.125151}
}

@article{Zollner_Fabian_2021,
  title = {Graphene on two-dimensional hexagonal BN, AlN, and GaN: Electronic, spin-orbit, and spin relaxation properties},
  author = {Zollner, Klaus and Cummings, Aron W. and Roche, Stephan and Fabian, Jaroslav},
  journal = {Phys. Rev. B},
  volume = {103},
  issue = {7},
  pages = {075129},
  numpages = {16},
  year = {2021},
  month = {Feb},
  publisher = {American Physical Society},
  doi = {10.1103/PhysRevB.103.075129},
  url = {https://link.aps.org/doi/10.1103/PhysRevB.103.075129}
}

@article{Sanchez_Stauber_2024,
  title = {Nematic versus Kekul\'e Phases in Twisted Bilayer Graphene under Hydrostatic Pressure},
  author = {S\'anchez S\'anchez, Miguel and D\'{\i}az, Israel and Gonz\'alez, Jos\'e and Stauber, Tobias},
  journal = {Phys. Rev. Lett.},
  volume = {133},
  issue = {26},
  pages = {266603},
  numpages = {7},
  year = {2024},
  month = {Dec},
  publisher = {American Physical Society},
  doi = {10.1103/PhysRevLett.133.266603},
  url = {https://link.aps.org/doi/10.1103/PhysRevLett.133.266603}
}
\end{document}


\graphicspath{}

\renewcommand{\figurename}{Supplementary Figure}

\title{Supplementary materials - Magnetism and hidden quantum geometry in charge neutral twisted trilayer graphene}

\author{Alina Wania Rodrigues} \thanks{awaniaro@uottawa.ca}
\affiliation{Department of Physics, University of Ottawa, Ottawa, Ontario, K1N 6N5, Canada}

\author{Maciej Bieniek} \thanks{maciej.bieniek@pwr.edu.pl}
\affiliation{Institute of Theoretical Physics, Wroc\l aw University of Science and Technology, Wybrze\.ze Wyspia\'nskiego 27, 50-370 Wroc\l aw, Poland}

\author{Daniel Miravet} \thanks{daniel.miravet@uottawa.ca}
\affiliation{Department of Physics, University of Ottawa, Ottawa, Ontario, K1N 6N5, Canada}

\author{Pawe\l\ Hawrylak} \thanks{pawel.hawrylak@uottawa.ca}
\affiliation{Department of Physics, University of Ottawa, Ottawa, Ontario, K1N 6N5, Canada}

\date{\today}

\maketitle

\begin{figure}\
\includegraphics[width=0.7\textwidth]{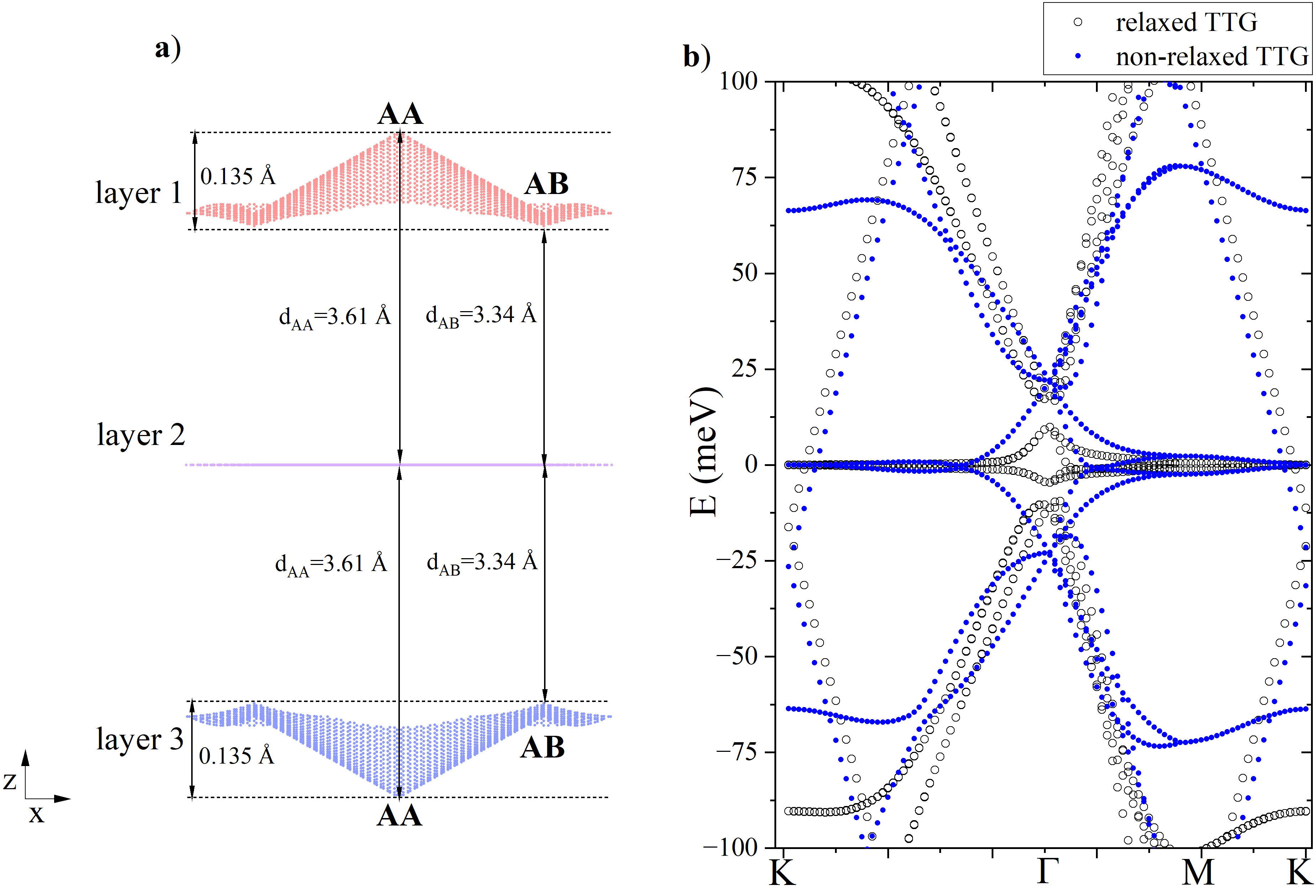}\
\caption{\textbf{Relaxation details.} (a) Schematic side view of relaxed TTG layers. The interlayer distance varies across the moir\'e unit cell depending on the local stacking. The middle layer remains rigid and top/bottom layers are adjusted. The layers are closest in the AB-stacked regions, where the interlayer spacing is \( d_{\rm AB} = d_{\rm min} = 3.34 \, \text{\AA} \). The largest separation occurs in the AA-stacked regions, where \( d_{\rm AA} = d_{\rm max} = 3.61 \, \text{\AA} \). (b) Band structure of non-relaxed and relaxed TTG denoted by blue dots and black circles, respectively. No gap between the flat and remote bands is observed for non-relaxed structure. Relaxation leads to flattening of the flat band, renormalization of the remote bands, and gaps opening between the flat and the remote bands an the $\Gamma$ point.
} \
\label{SM-relaxation}
\end{figure}\

\begin{figure}\
\includegraphics[width=1\textwidth]{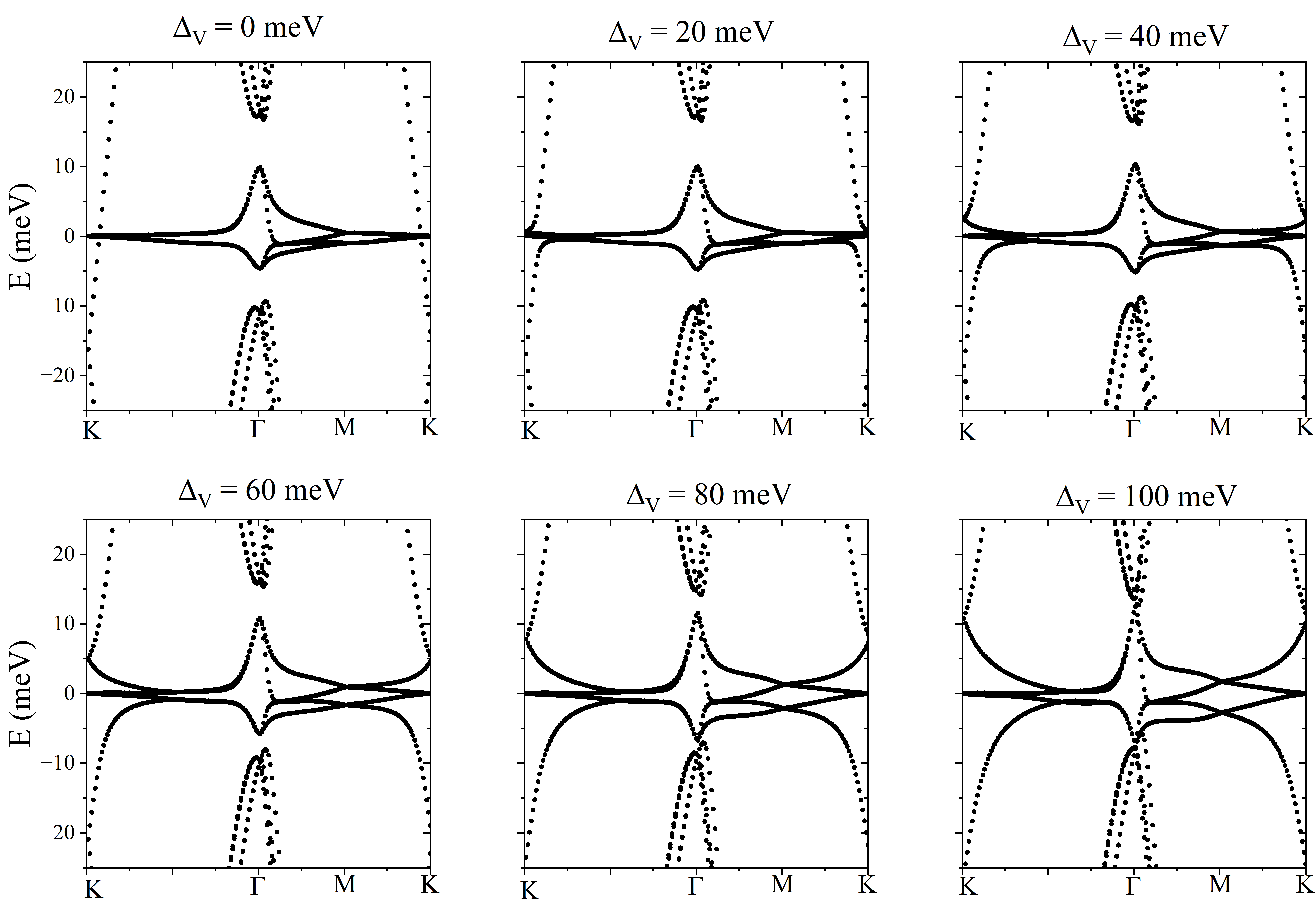}\
\caption{\textbf{Effect of electric field.} Renormalization of the non-interacting bandstructure as a function of the applied electric field $\Delta_V$. We show here an evolution of the band gaps opening within the flat band and splitting off of the Dirac cones. For $\Delta_V$ that's large enough ($\geq 100$ meV) the gap between the flat band and the remote bands closes at the $\Gamma$ point.} \
\label{SM-electric_field}
\end{figure}\

\begin{figure}\
\includegraphics[width=0.85\textwidth]{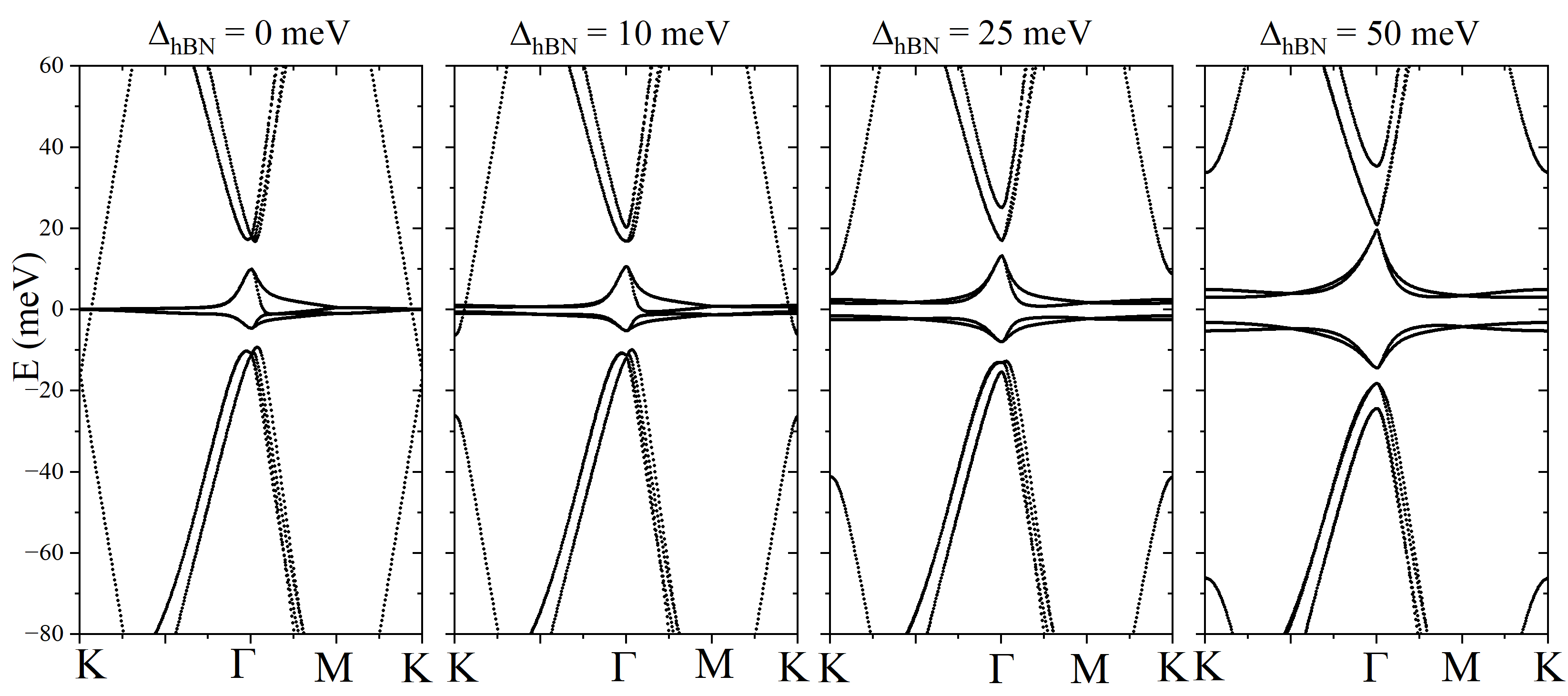}\
\caption{\textbf{Effect of hBN.} Analogous to Sup. Fig. 2 evolution of bandstructure as a function of the strength of the staggered potential $\Delta_{\rm hBN}$ induced by an hBN substrate. Increase of $\Delta_{\rm hBN}$ leads to a gap opening within the flat band and between the Dirac cones at $K$-points.} \
\label{SM-hBN}
\end{figure}\

\begin{figure}\
\includegraphics[width=1\textwidth]{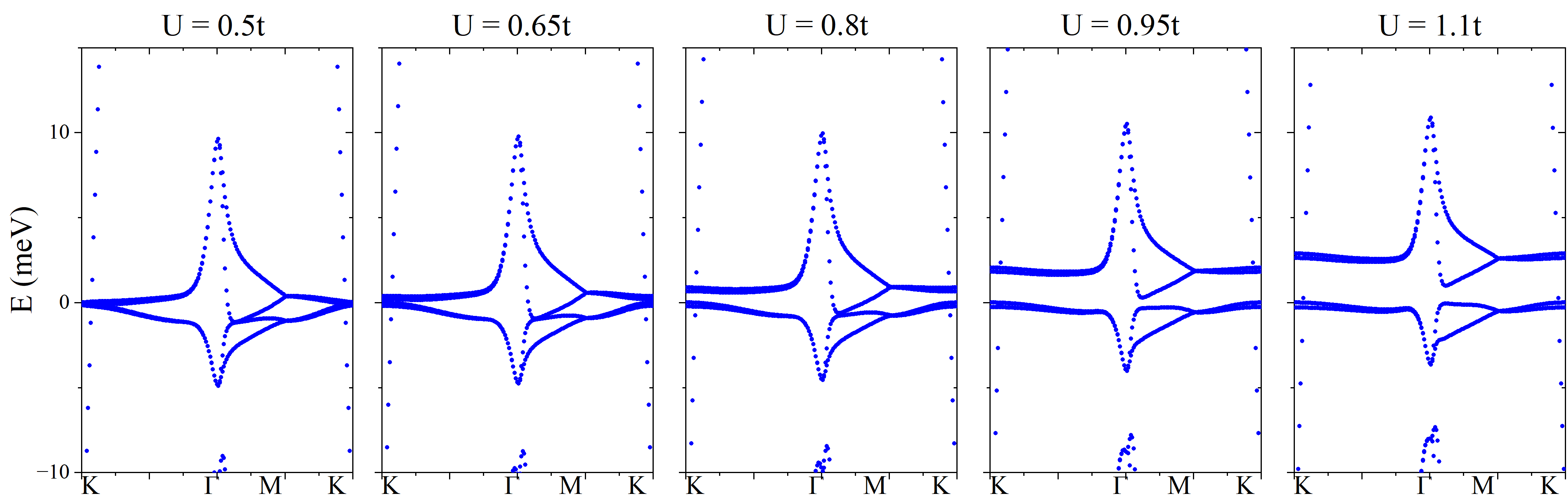}\
\caption{\textbf{Effect of interaction.} Renormalization of the mean-field band structure as a function of the interaction strength $U$, with $\Delta_V = 0$ and $\Delta_{\rm hBN}=0$. As $U$ increases, Mott-like gap within the flat band opens and the bands along $K-\Gamma$ and $M-K$ flatten. We observe no gap opening within high-velocity Dirac cones, therefore no global gap is generated.} \
\label{SM-hBN}
\end{figure}\

\begin{figure}\
\includegraphics[width=1\textwidth]{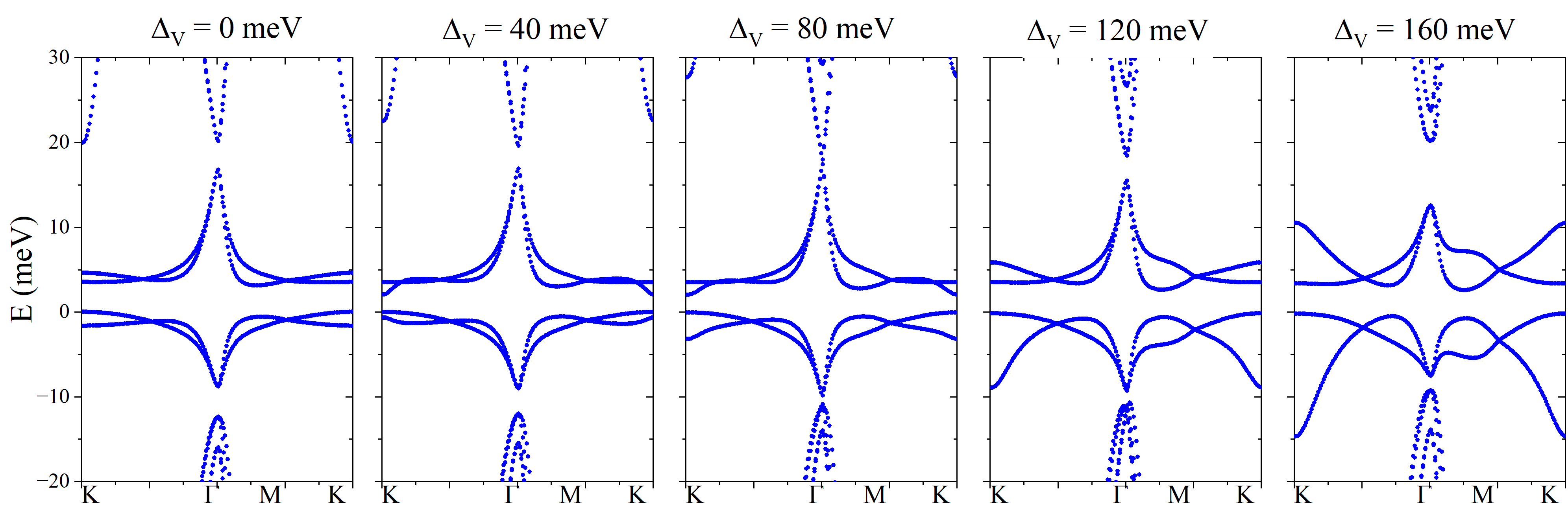}\
\caption{\textbf{Effect of electric field on a mean-field band structure} Spectrum of interacting TTG with hBN encapsulation as a function of applied electric field $\Delta_V$. Hubbard parameter $U=0.95t$, $\Delta_{\rm hBN}=50$ meV. As the $\Delta_V$ increases, the gap between the flat and remote bands decreases, and finally closes for $\Delta_V=80$meV. However, further increase of the electric field leads to the gap opening again.}\
\label{SM-U}
\end{figure}\

\begin{figure}\
\includegraphics[width=0.8\textwidth]{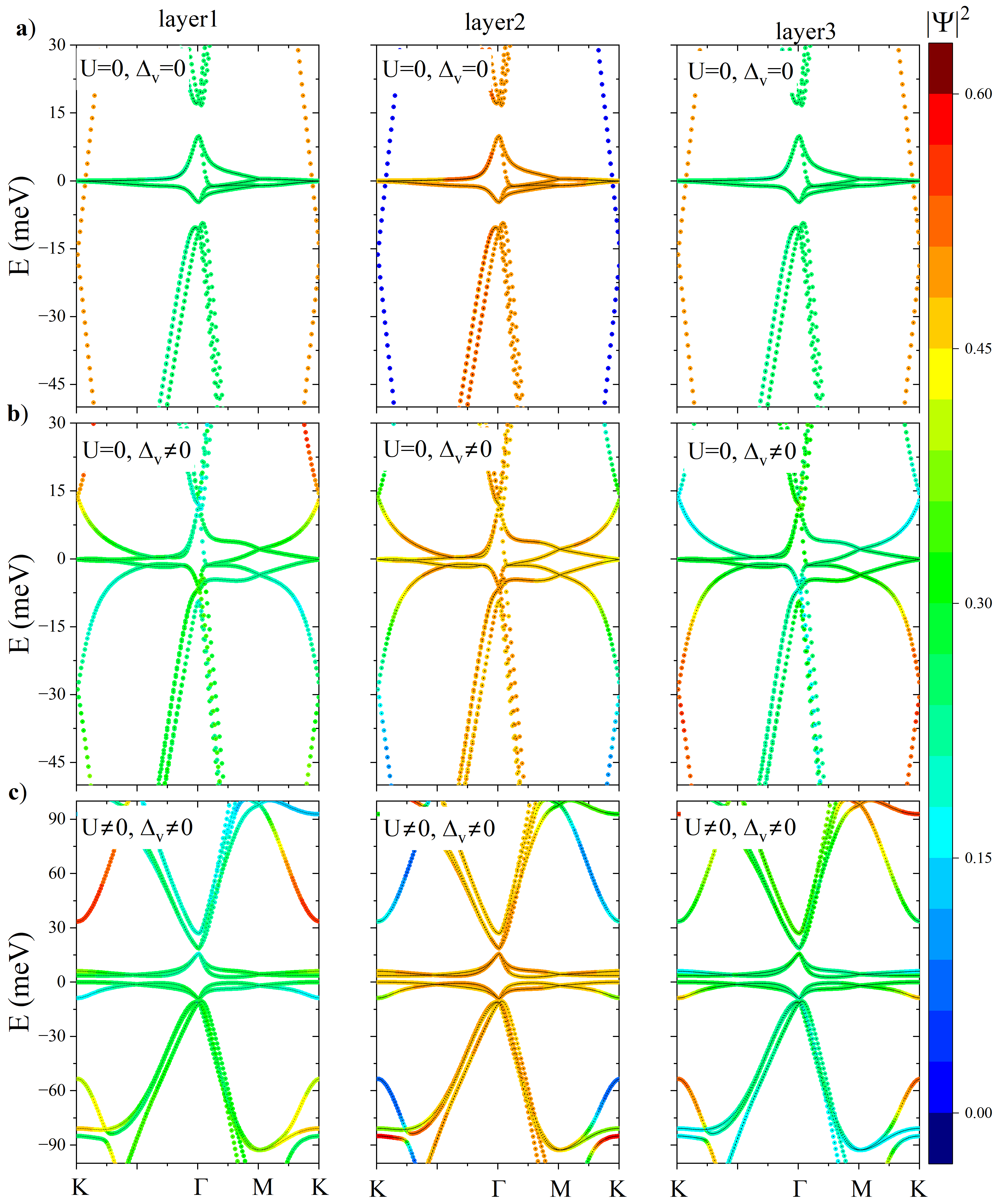}\
\caption{\textbf{Layer-resolved wave function localization of Bloch states.} a) Non-interacting TTG, (b) non-interacting TTG with applied electric field, (c) interacting TTG with applied electric field. The color scale shows the localization of the wave function $|\Psi|^2$ in layer 1 (left panels), layer 2 (middle panels) and layer 3 (right panels). In the case of (a) mirror symmetry is preserved and the wave function is equally distributed between the top and bottom layer. The wave functions of the Dirac cones are localized solely on the top and bottom layers, while the flat band's wave functions are localized $50\%$ on the middle layer and $25\%$ on the top and bottom. (b) Applying electric field breaks the mirror symmetry and the wave function distribution is no longer equal between the top and bottom layer. (c) Including interactions further renormalizes this distribution, however one can still notice the clear trend of majority of the Dirac cone's wave function being localized in the top and bottom layers and the flat band wave function being localized predominantly in the middle layer.} \
\label{SM-wf1}
\end{figure}\

\begin{figure}\
\includegraphics[width=0.8\textwidth]{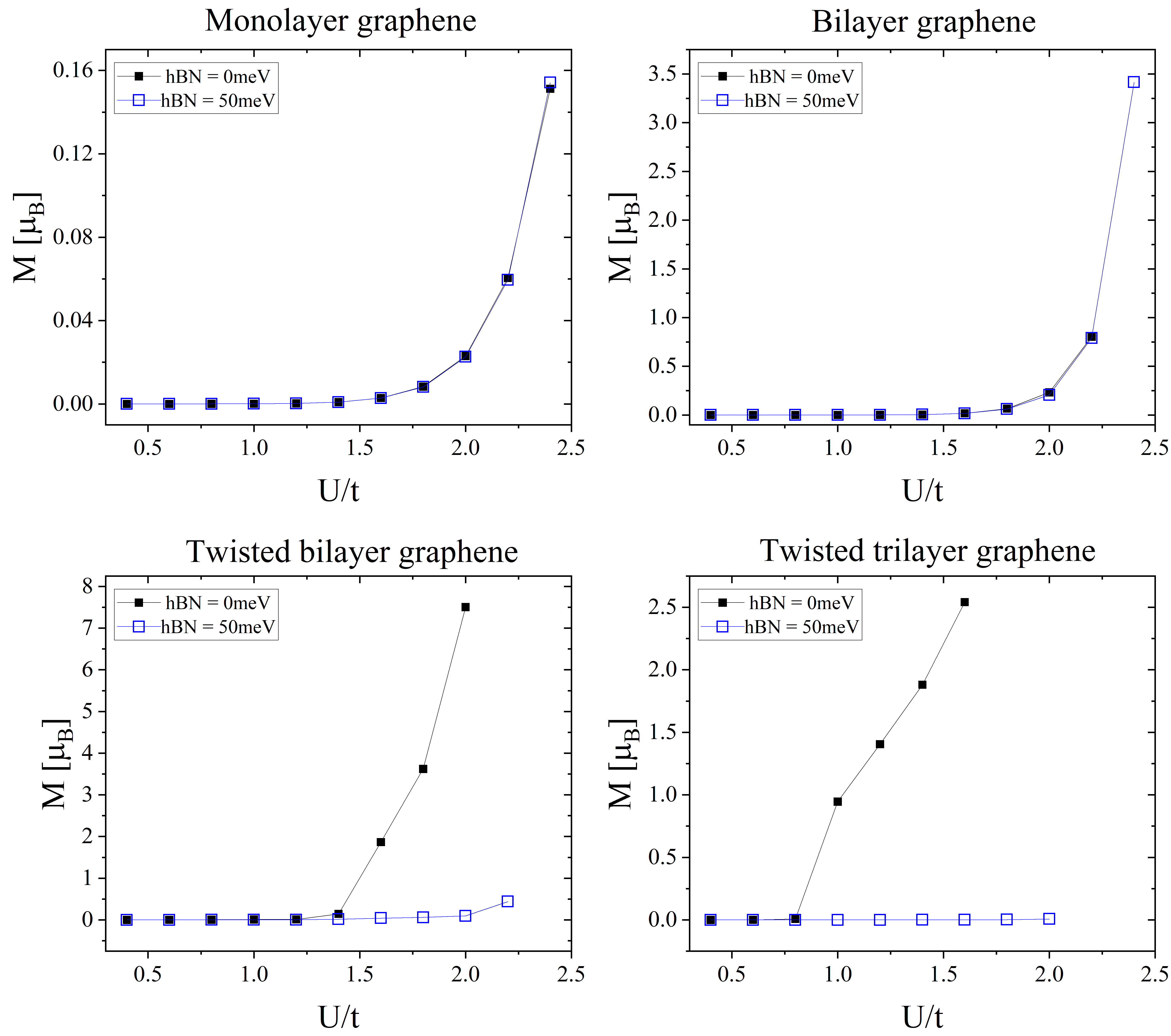}\
\caption{\textbf{AF transition in various graphene-based systems} with (blue empty squares) and without (black filled squares) the hBN substrate. Total absolute magnetization of different graphene systems in the units of Bohr magneton as a function of the interaction $U$ scaled by the nearest-neighbor hopping $t$. In twisted bilayer and trilayer graphene systems, the inclusion of a staggered potential has a similar effect - it delays the transition from the paramagnetic to the antiferromagnetic phase. In contrast, for monolayer and untwisted bilayer graphene, the impact of the staggered potential is negligible. This is because the para-/antiferromagnetic transition in these systems occurs at much larger interaction strengths U, resulting in a high $U/\Delta_{hBN}$ ratio ($\sim$ 40). Consequently, interaction effects dominate over those induced by the staggered potential.} \
\label{SM-AF}
\end{figure}\

\bibliography{bibliography_Dec2024}